\documentclass{osa-article}
\DeclareUnicodeCharacter{0306}{\v}

\journal{osajournal}

\articletype{Research Article}

\begin{document}

\title{Phase noise mitigation by a realistic optical parametric oscillator}

\author{Michele~N.~Notarnicola\authormark{1}, Marco~G.~Genoni\authormark{1,2}, Simone~Cialdi\authormark{1,2}, Matteo~G.~A.~Paris\authormark{1,2}, and Stefano~Olivares\authormark{1,2,*}}

\address{\authormark{1} Dipartimento di Fisica “Aldo Pontremoli,” Università degli Studi di Milano, I-20133 Milano, Italia\\
\authormark{2} Istituto Nazionale di Fisica Nucleare, Sezione di Milano, I-20133 Milano, Italia}
\email{\authormark{*} stefano.olivares@fisica.unimi.it} 

\begin{abstract}
We address the exploitation of an optical parametric oscillator (OPO) in the task of mitigating, at least partially, phase noise produced by phase diffusion. In particular, we analyze two scenarios where phase diffusion is typically present. The first one is the measurement of the phase of a noisy optical field, while the second involves a quantum estimation scheme of a phase shift imposed on a noisy probe. In both cases, we prove that an OPO may lead to a partial or full compensation of the noise.
\end{abstract}

\section{Introduction}
The phase of an optical field is a fundamental degree of freedom for several applications in quantum sensing and quantum communication. Quantum interferometry is exploited in high-precision measurements to detect fine perturbations through phase shifts \cite{Caves1981, demkowicz2009, sparaciari2015} and continuous-variable communication protocols are often based on \textit{phase shift keying} (PSK) \cite{Kazovsky2006, Olivares2013, mondin15}. However, the optical phase cannot be described as an observable in a strict sense \cite{Susskind, Louisell, Paris-1994, Lalovic} and this result makes it challenging to provide a detailed description of all the strategies involving the phase.

In practical contexts the phase of a field is often affected by phase noise especially due to phase diffusion.
Indeed, the presence of phase diffusion may lead to a partial or complete loss of all the advantages of quantum measurements. 
In the quantum optics scenario, phase diffusion efficiently describes the noisy propagation of quantum light through optical fibers \cite{Ezra2008}, and its effect has been thoroughly investigated on phase estimation and quantum communication protocols \cite{Qubit2, Qubit1,Genoni,Genoni2012,Trapani2015,Jarzyna2016,Bina,DiMario2019}. More generally, its detrimental effects have been also investigated for different physical platforms, such as Bose-Einstein condensates \cite{Bose1, Bose2} and Bose-Josephson junctions \cite{Josephson}.

Focusing our attention on quantum optical systems, one of the possible resources to counteract phase noise is provided by a phase-sensitive amplifier, such as a \textit{optical parametric oscillator} (OPO). More precisely, an OPO is not expected to be useful to improve measurements, i.e. to build receivers, because the induced phase shift would lead to a modification of the phase outcome. On the contrary, an OPO may be very effective in order to improve the properties of the probe state if affected by phase noise.

In this work, we address the possibility of using an OPO to compensate, at least partially, the detriments of phase noise. In particular, we present a theoretical model approaching the task of removing phase noise of a laser. Our method could be effective to reduce the laser phase noise at high frequency (between 100 kHz and 10 MHz). At these frequencies it is hard to distinguish between amplitude and phase noise to act on the phase noise, but our technique acts directly on the phase and it is not influenced by the amplitude noise. Hints that squeezing could help in this scenario have already been shown in \cite{Cialdi,Carrara}. Here we discuss in more detail a realistic experimental implementation of a squeezing operation via an OPO, taking into account the most relevant experimental details. 
Moreover, we apply the scenario described above to two different cases where the optical phase is exploited.
The first one is a pure \textit{quantum optical} context and regards the measurement of the phase of a quantum state of radiation. On the contrary, the second case consists in a \textit{quantum estimation} scheme where information is encoded on a phase shift. In both the cases we consider single mode radiation prepared in a coherent state $|\alpha\rangle$, $\alpha \in \mathbb{R}_+$ undergoing phase diffusion. Just after dephasing, we introduce an OPO and discuss if its exploitation may lead to noise mitigation. 

The paper is structured as follows. First of all, in Sec.~\ref{sec:OPO} we present a theoretical model for the OPO, described in the Schr\"odinger picture, allowing to work directly at the level of quantum states. Bearing that in mind, we apply it to the two cases in exam. In Sec. \ref{sec: OpticalPhase} we address the measurement of an optical phase and state in which conditions the OPO is able to counteract phase noise while Sec.~\ref{sec: PhaseShift} is dedicated to describe the estimation scheme with the tools of quantum estimation theory and to find out the optimal measurement to detect the value of the encoded phase shift with the highest possible precision. We close our investigation drawing some concluding remarks in Sec.~\ref{s:concl}.

\section{A block-diagram model for the OPO}\label{sec:OPO}
\begin{figure}
\centering
\includegraphics[width=.6\textwidth]{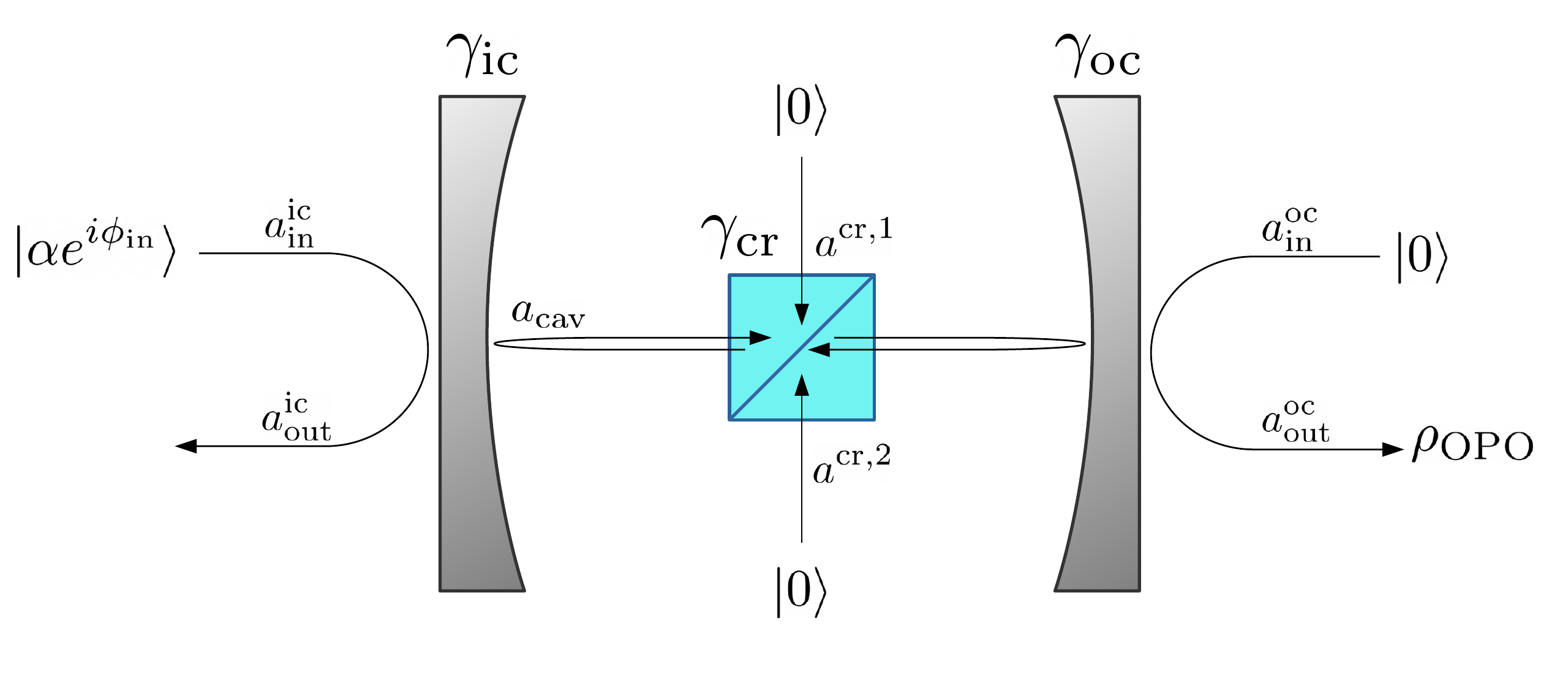}\\
\includegraphics[width=.6\textwidth]{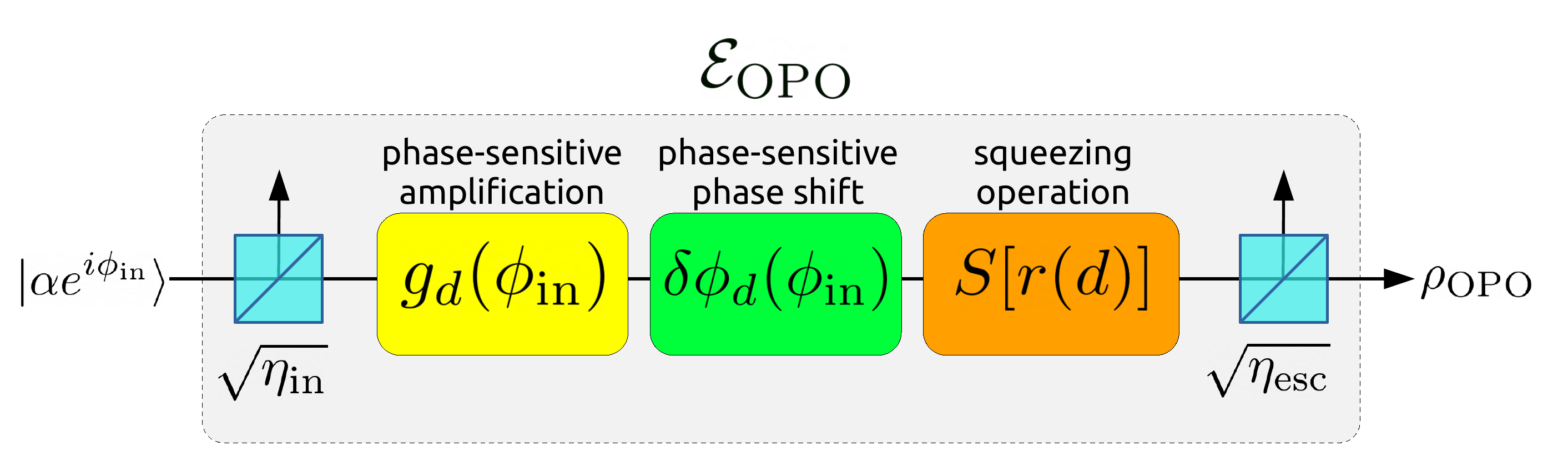}
\caption{Top: schematic diagram of an OPO in the input-output description with a coherent state $|\alpha e^{i \phi_{\rm{in}}}\rangle$ entering the input coupler. Bottom: Block scheme. The dynamics of the OPO can be described as the subsequent application of phase-sensitive amplification $g_d(\phi_{\rm{in}})$, phase-sensitive phase shift $\delta\phi_d(\phi_{\rm{in}})$ and squeezing $S[r(d)]$.}\label{fig:OPO}
\end{figure} 

A traditional description of the OPO is obtained by exploiting the input-output formalism \cite{aguideto}. As depicted in the top panel of Fig.~\ref{fig:OPO}, the dynamics of the OPO is characterized by the input and output modes $a_{\rm{in/out}}^{\rm{ic/oc}}$ associated with the input and output coupler, the input modes associated with the crystal losses $a^{\rm{cr,1/cr,2}}$ and the coherent evolution of the cavity mode $a_{\rm{cav}}$ generated by the squeezing Hamiltonian $H_s=i \frac12 g (a^{\dagger 2}_{\rm{cav}}- a^2_{\rm{cav}})$, where we have assumed to fix the laser pump in order to amplify the quadrature $q= (a_{\rm{cav}}+a^\dagger_{\rm{cav}})/\sqrt{2}$ of the field inside the cavity. The input-output operators satisfy the canonical commutation relations (CCR) $[a_{\rm{in/out}}^{\rm{ic/oc}}, a_{\rm{in/out}}^{ \rm{ic/oc} \dagger}]=[a^{\rm{cr,1/cr,2}}, a^{\mathrm{cr,1/cr,2} \dagger}]=1$, while the cavity mode evolves such that $[a_{\rm{cav}}(t),a^\dagger_{ \rm{cav}}(t')]= \delta(t-t')$ \cite{aguideto}.
We now introduce the parameters $\eta_{\rm{in}}$ and $\eta_{\rm{esc}}$ and the squeezing parameter $d$. The input and output parameters depend on the loss rates at both the input coupler ($\gamma_{\rm{ic}}$) and the output coupler ($\gamma_{\rm{oc}}$) and the rate of internal losses of the crystal ($\gamma_{\rm{cr}}$), see Fig.~\ref{fig:OPO}. We have 
\begin{equation}
\eta_{\rm{in}}= \gamma_{\rm{ic}}/\gamma, \quad \eta_{\rm{esc}}= \gamma_{\rm{oc}}/\gamma,
\end{equation}
where $\gamma = \gamma_{\rm{ic}}+\gamma_{\rm{oc}}+ 2 \gamma_{\rm{cr}}$. The squeezing parameter $d$ is proportional to the coupling of the squeezing Hamiltonian and reads $d= g/\gamma$ and the stability condition of the OPO imposes $d<1$ \cite{aguideto}.

By considering a roundtrip of the cavity of duration $\tau$, defining $\tilde{a}_{\rm{cav}}= \sqrt{\tau} a_{\rm{cav}}$ and in the presence of high reflectivity mirrors, the Langevin equation for the cavity mode and its boundary condition read \cite{aguideto}
\begin{subequations}\label{eq:Langevin}
\begin{align}
 \frac{d \tilde{a}_{\rm{cav}}}{dt}&=  g \tilde{a}_{\rm{cav}}^{\dagger}(t) -\gamma \tilde{a}_{\rm{cav}}(t) +\sqrt{2 \gamma_{\mathrm{ic}}} \ a^{\mathrm{ic}}_{\mathrm{in}} \notag\\
 &\hspace{1cm}+ \sqrt{2 \gamma_{\mathrm{oc}}} \ a^{\mathrm{oc}}_{\mathrm{in}} +\sqrt{2 \gamma_{\mathrm{cr}}} (a^{\mathrm{cr, 1}}+a^{\mathrm{cr,2}}) , \\[1ex]
a^{\mathrm{ic/oc}}_{\mathrm{out}}&= -a^{\mathrm{ic/oc}}_{\mathrm{in}} + \sqrt{2 \gamma_{\mathrm{ic/oc}}} \ \tilde{a}_{\rm{cav}}.
\end{align}
\end{subequations}
Considering the device in the stationary regime, Eqs.~(\ref{eq:Langevin}) can be exploited to express the output mode in function of the input ones as
\begin{align}
a^{\rm{oc}}_{\rm{out}} &= d \ (a^{\rm{oc} \dagger}_{\rm{out}}+ a^{\rm{oc} \dagger}_{\rm{in}}) +  (\eta_{\rm{esc}}-\eta_{\rm{in}}-2\eta_{\rm{cr}}) a^{\rm{oc}}_{\rm{in}} +\notag\\[1ex]
& \hspace{1cm}+2 \sqrt{\eta_{\rm{in}}\eta_{\rm{esc}}} a^{\rm{ic}}_{\rm{in}} + 2 \sqrt{\eta_{\rm{cr}}\eta_{\rm{esc}}} (a^{\rm{cr, 1}}+a^{\rm{cr,2}}),\label{eq: Stationary}
\end{align}
with $\eta_{\rm{cr}}= \gamma_{\rm{cr}}/\gamma$. The last equation is not in closed form, therefore it is convenient to pass to quadratures
\begin{equation}
q_{\rm{out}}= \frac{a^{\rm{oc}}_{\rm{out}}+ a^{\rm{oc} \dagger}_{\rm{out}}}{\sqrt{2}},\quad
p_{\rm{out}}= \frac{a^{\rm{oc}}_{\rm{out}}- a^{\rm{oc} \dagger}_{\rm{out}}}{\sqrt{2}\ i}.
\end{equation}

For an initial coherent state at the input coupler $|\alpha e^{i \phi_{\rm{in}}}\rangle$, $\alpha \in \mathbb{R}_+$, and the vacuum in all other ports, the final state at the output coupler $\rho_{\rm{OPO}}$ is such that
\begin{subequations}\label{eq:moments}
\begin{align}
\bigl\{ \langle q_{\rm{out}} \rangle , \langle p_{\rm{out}} \rangle \bigr\}&=
\sqrt{2} \bigl\{ \tilde{\alpha}_q \cos\phi_{\rm{in}} , \tilde{\alpha}_p \sin\phi_{\rm{in}} \bigr\} ,\\[1ex]
\mathrm{Var}[q_{\rm{out}}]  &=
\frac{1}{2} \biggl[1+ \eta_{\rm{esc}} \frac{4d}{(1 - d)^2} \biggr] \equiv \Sigma^2_{q} \,,\\[1ex]
\mathrm{Var}[p_{\rm{out}}]  &=
\frac{1}{2} \biggl[1 - \eta_{\rm{esc}} \frac{4d}{(1+ d)^2} \biggr] \equiv \Sigma^2_{p} \,,
\label{eq:VarOPO}
\end{align}
\end{subequations}
with 
\begin{align}\label{eq:tildealpha}
\tilde{\alpha}_q = 2 \frac{ \sqrt{ \eta_{\rm{in}} \eta_{\rm{esc}}}\ \alpha }{1 - d },\quad
\tilde{\alpha}_p = 2 \frac{ \sqrt{ \eta_{\rm{in}} \eta_{\rm{esc}}}\ \alpha}{1+ d}.
\end{align}
In Eq.~(\ref{eq:VarOPO}) we clearly have $\Sigma^2_p < 1/2$, showing squeezing in the $p_{\rm{out}}$ quadrature. Moreover, two other effects induced by the OPO are a phase-sensitive amplification and a phase-sensitive phase shift. By computing expectation values on Eq.~(\ref{eq: Stationary}) with the same input states, we get
\begin{equation}
\langle a^{\rm{oc}}_{\rm{out}} \rangle
= 2\sqrt{\eta_{\rm{in}} \eta_{\rm{esc}}} \ \alpha_{\rm{out}} \ e^{i \phi_{\rm{out}}},
\end{equation}
where
\begin{subequations}
\begin{align}
&\alpha_{\rm{out}} = \frac{ \alpha \sqrt{1+ 2 d \cos(2 \phi_{\rm{in}})+d^2}}{1-d^2}, \\[1ex]
&\tan \phi_{\rm{out}} = \frac{1+d}{1-d} \ \tan \phi_{\rm{in}} .
\end{align}
\end{subequations}

An equivalent description of the OPO may be obtained working in the Schr\"odinger picture, through the block scheme sketched in the bottom panel of Fig.~\ref{fig:OPO}. We consider a single mode of radiation $a$ entering the OPO, $[a, a^\dagger]= 1$. Then the block scheme consists of the sequential application of unitary operations associated with all the transformations produced by the OPO, that is beam splitters of transmissivity $\eta_{\rm{in/esc}}$ for the input and output couplers, respectively, phase-sensitive amplification $g_d(\phi_{\rm{in}})$, phase-sensitive phase shift $\delta \phi_d(\phi_{\rm{in}})$ and squeezing $S[r(d)]= \exp\{\frac12 r(d)\ [a^{\dagger 2} - a^2]\}$. In order to obtain an output state with the expectations given in Eqs.~(\ref{eq:moments}), we should set
\begin{subequations}
\begin{align}
g_d(\phi_{\rm{in}}) &= 2 \frac{\sqrt{1-2d\cos(2\phi_{\rm{in}})+d^2}}{1-d^2} ,\\[1ex]
\delta \phi_d (\phi_{\rm{in}}) &= \arctan \biggl( \frac{1+d}{1-d} \tan \phi_{\rm{in}}\biggr) - \phi_{\rm{in}} ,\\[1ex]
\exp[r(d)] &= \frac{1+d}{1-d}.\label{eq: squeezing_r}
\end{align}
\end{subequations}

The block scheme defines a quantum CP-map $\mathcal{E}_{\rm{OPO}}$ such that $\mathcal{E}_{\rm{OPO}}(|\alpha e^{i \phi_{\rm{in}}}\rangle \langle \alpha e^{i \phi_{\rm{in}}}|) = \rho_{\rm{OPO}}$ and 
this approach proves to be equivalent to the input-output one.
Moreover, since the block scheme involves the subsequent application of unitary operations associated with linear or bilinear Hamiltionians \cite{OlivaresGauss}, we conclude that $\rho_{\rm{OPO}}$ is a Gaussian state and $\mathcal{E}_{\rm{OPO}}$ is a Gaussian CP-map. Therefore, prime and second moments suffice for a full description of the output state.

\section{Case I -- Phase measurement using a noisy probe}\label{sec: OpticalPhase}

\begin{figure}[ht!]
\centering
\includegraphics[width=.75\textwidth]{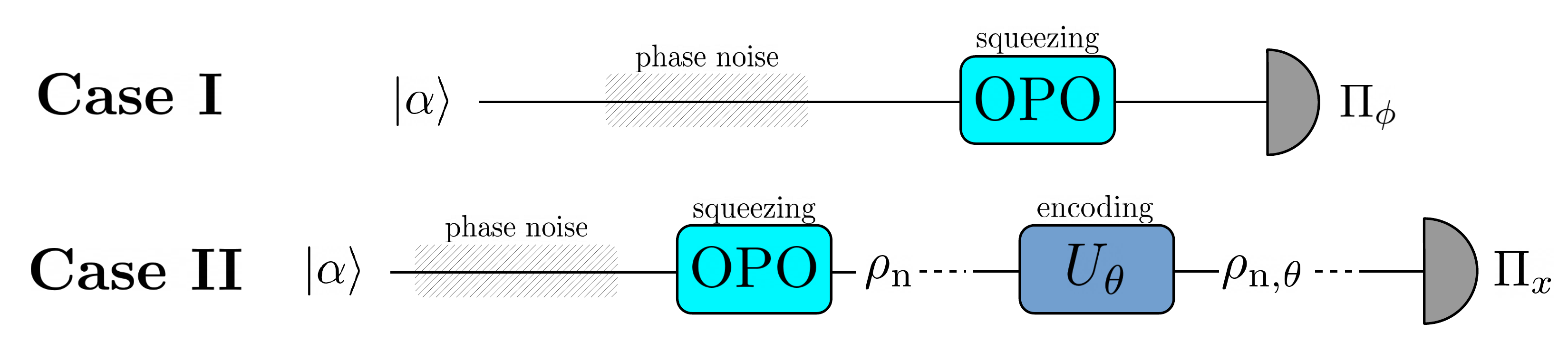}
\caption{Schematic diagram of the two scenarios discussed in this paper.}\label{fig: Cases}
\end{figure}

Let us now address the case I of Fig.~\ref{fig: Cases}. We consider as probe a coherent state $\rho_0= |\alpha\rangle \langle \alpha|$, $\alpha \in \mathbb{R}_+$, assuming, without loss of generality, the average value of its phase to be fixed. Indeed, it is typically possible to control the average phase by employing a suitable feedback protocol \cite{PhysRevA.54.4495,Bina}. The coherent seed undergoes phase noise (Fig.~\ref{fig: Cases}), whose overall effect is the application of a random phase shift Gaussian-distributed \cite{Cialdi, Carrara, Genoni}. That is
\begin{equation}\label{eq:PhaseDiff}
\rho_D= \mathcal{E}_{\sigma} (|\alpha\rangle \langle \alpha|)= \int_{\mathbb{R}} d\psi\
\frac{e^{-\psi^2/(2 \sigma^2)}}{\sqrt{2 \pi \sigma^2}}\
U_\psi  |\alpha\rangle \langle \alpha|  U^\dagger_\psi,
\end{equation}
where $U_\psi= e^{-i \psi a^\dagger a}$ is a phase shift operation and $\sigma$ is the amplitude of the noise.
Then, following the case I of Fig.~\ref{fig: Cases}, we let the dephased state pass through an OPO, described by the map $\mathcal{E}_{\rm{OPO}}$, obtaining the final state $\rho_{\rm{out}}= \mathcal{E}_{\rm{OPO}} \bigl(\mathcal{E}_\sigma (|\alpha\rangle \langle \alpha|) \bigr)$. The task is to find a feasible strategy to measure the optical phase and to state whether the OPO is able to compensate the effects of phase noise by comparing the results obtained with states $\rho_D$ and $\rho_{\rm{out}}$.

\begin{figure}
\centering
\includegraphics[width=.4\textwidth]{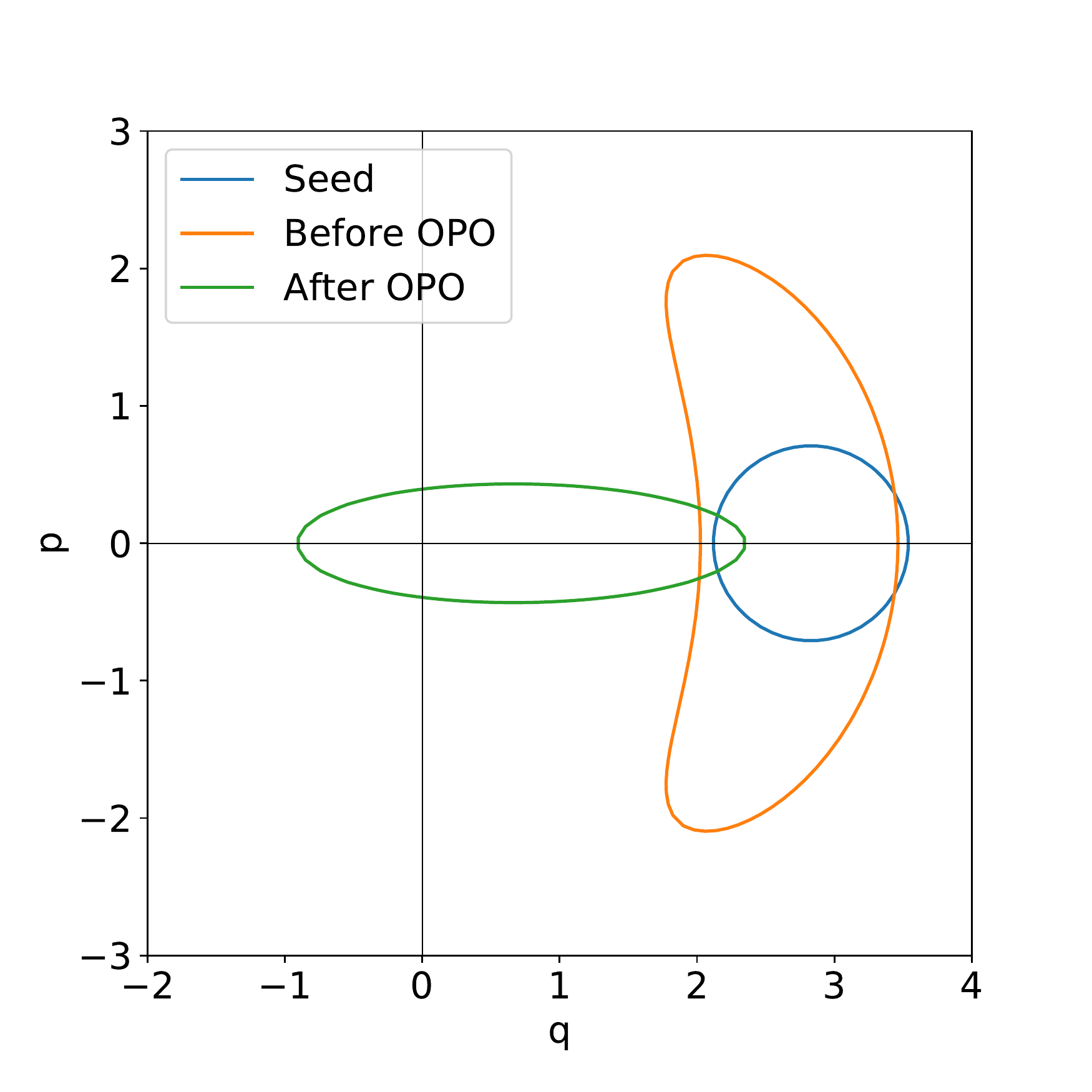} \\ 
\includegraphics[width=.6\textwidth]{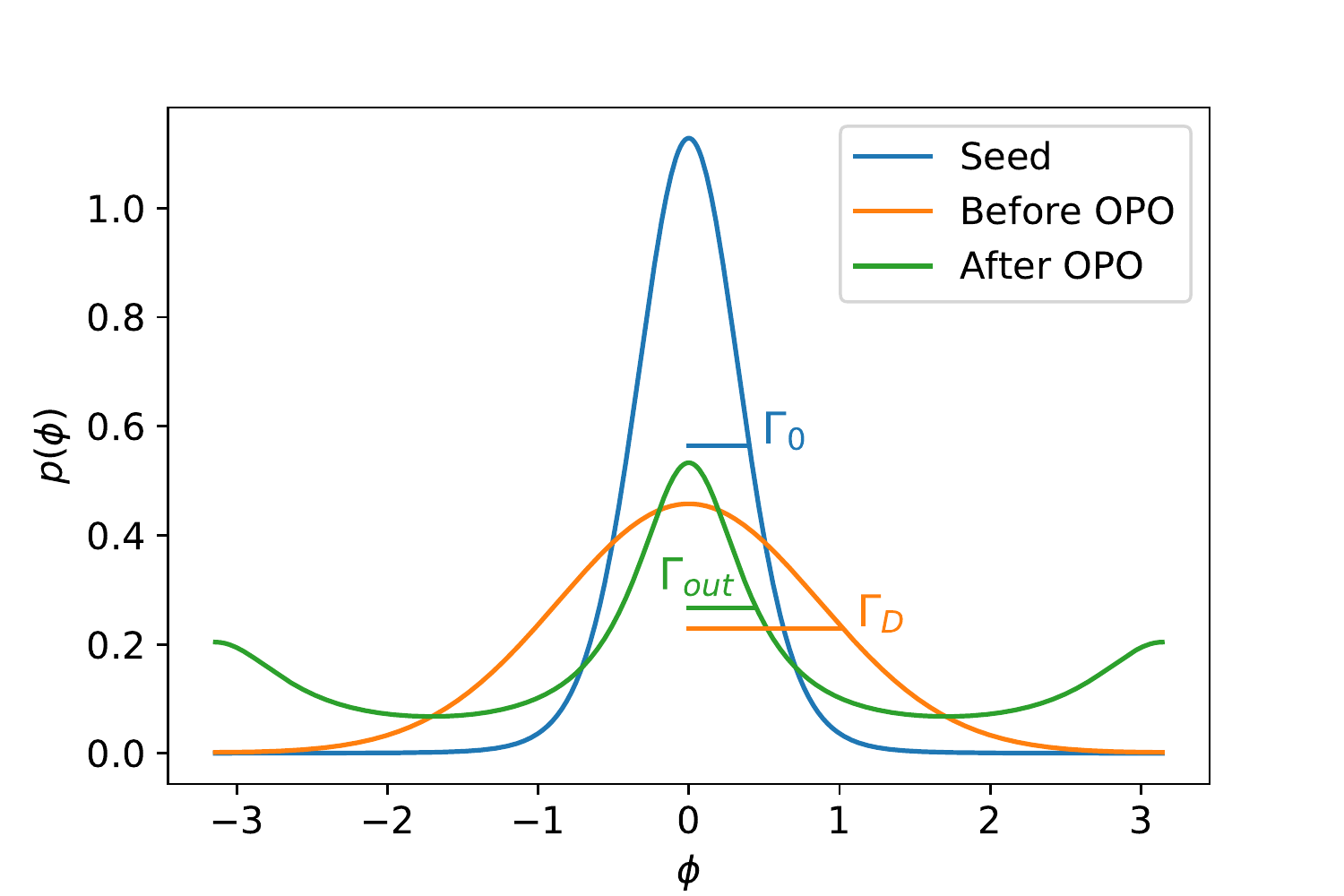}
\caption{Top: phase space representation of $\rho_0$, $\rho_D$ and $\rho_{\rm{out}}$ where we plot the level curves at the level of the standard deviation. Bottom: phase distributions $p_0(\phi)$, $p_D(\phi)$, $p_{\rm{out}}(\phi)$. We set $\alpha=2$, $\sigma=\pi/4$, $d=0.4$ and we used the realistic parameters $\eta_{\rm{in}}=0.01$, $\eta_{\rm{esc}}=0.93$.}\label{fig:phasespace}
\end{figure}

Fig. \ref{fig:phasespace} shows the effects of such transformations at the level of quantum states. In the phase space dephasing causes a in-homogeneous spread of the coherent state which makes state $\rho_D$ not Gaussian anymore. Then, the action of the OPO squeezes quadrature $p$ at the expense of magnifying the variance of $q$.

To give an estimate of the phases of $\rho_D$ and $\rho_{\rm{out}}$ and assess whether the OPO can \textit{squeeze} the noise, we will consider two different approaches. First of all, we perform a \textit{direct measurement} of the phase, exploiting a phase POVM. Then, we present an \textit{indirect measurement} procedure based on a post processing on the data of two separate homodyne detections, which we will prove to give the same \textit{qualitative} results in the regime of $\alpha\gg 1$. Actually, given this scenario, we have a priori information that the value of the phase is $0$. “However, the purpose of both strategies is not the estimation of the phase value, but rather to state whether or not the OPO is able to reduce a proper figure of merit representing the phase uncertainty. Furthermore, we want to underline that the goal of this section is not to make a direct comparison between the two measurements proposed. The basic idea is to show that a realistic OPO may be a convenient resource to mitigate the noise and that this convenience is guaranteed with different phase measurement strategies employed.

\subsection{Direct measurement of the phase}

\begin{figure}
\centering
\includegraphics[width=.5\textwidth]{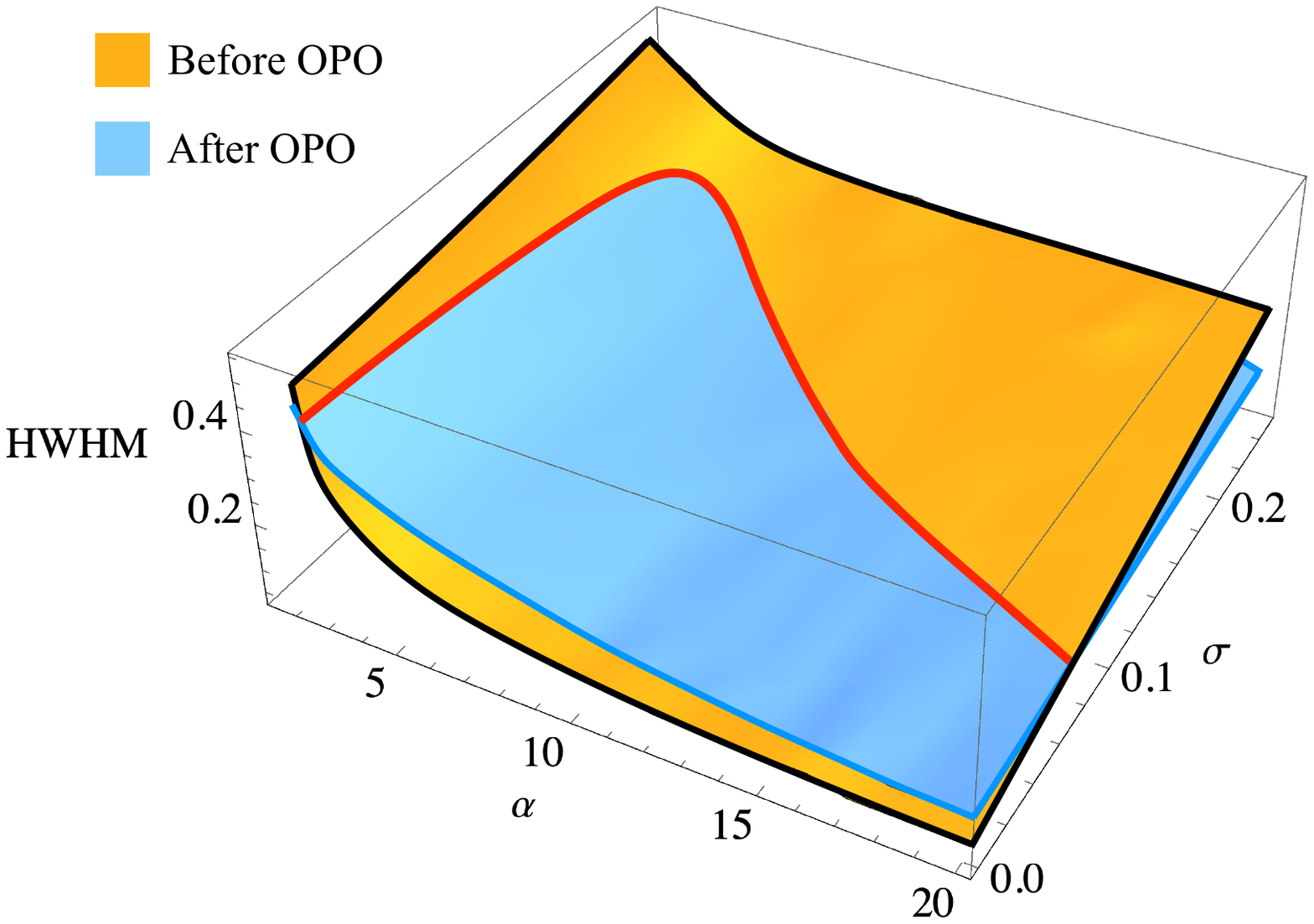} \\
\includegraphics[width=.6\textwidth]{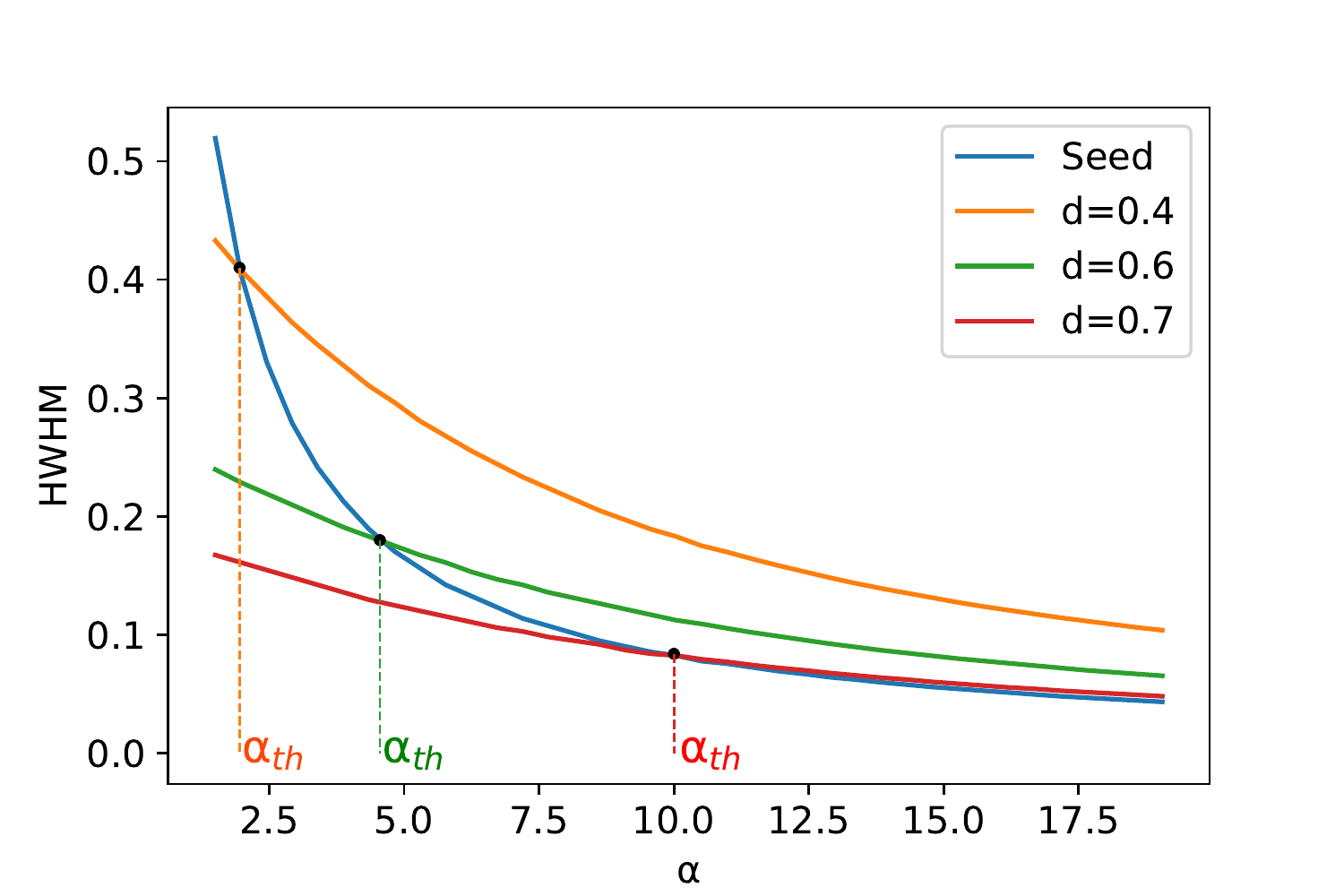}

\caption{Top: 3D plot of the dephased HWHM $\Gamma_D$ and the squeezed HWHM $\Gamma_{\rm{out}}$ as function of $\alpha$ and $\sigma$, with $d=0.4$. There exists a threshold signal $\alpha_{\rm{th}}$ such that for smaller $\alpha$, $\Gamma_{\rm{out}}<\Gamma_D$ for all $\sigma$, while for larger $\alpha$ the intersection between the two surfaces identifies the threshold noise $\sigma_{\rm{th}}$. Bottom: plot of the HWHMs $\Gamma_0$ of the seed and $\Gamma_S$ for different values of $d$ as functions of $\alpha$ in the absence of noise ($\sigma=0$). The intersection between $\Gamma_0$ and $\Gamma_S$ defines the threshold $\alpha_{\rm{th}}$, which appears to be an increasing function of $d$. We used the realistic parameters $\eta_{\rm{in}}=0.01$, $\eta_{\rm{esc}}=0.93$.}\label{fig:Phasediag}
\end{figure}

A convenient way to perform phase measurement is to employ a genuine phase POVM \cite{Paris-1994, Lalovic,PhysRevLett.67.1426}. Among all possible choices, here we consider a feasible POVM, implemented through a heterodyne detection, namely
\begin{equation}
\Pi_\phi = \frac{1}{\pi} \int_0^\infty d\zeta \ \zeta \ |\zeta e^{i \phi}\rangle\langle \zeta e^{i \phi}|,
\end{equation}
where $|\zeta e^{i \phi} \rangle$ is a coherent state. The corresponding phase probability for a given state $\rho$ reads  
\begin{align}
p(\phi)&= \mathrm{Tr}[\rho \Pi_\phi],\\[1ex]
&= \int_0^\infty d\zeta \ \zeta\ Q[\rho] (\zeta e^{i \phi}),
\end{align}
that is the marginal in phase of the Husimi $Q$-function $Q[\rho](z) = \langle z | \rho | z \rangle/\pi$, $z \in {\mathbb C}$.

Within this scenario, we have to compare the probabilities $p_0(\phi)$, $p_D(\phi)$ and $p_{\rm{out}}(\phi)$ associated with the initial state $\rho_0$, the dephased state $\rho_D$ and the output state $\rho_{\rm{out}}$, respectively, and we have to assess in which conditions the OPO leads to a reduction of the width of the distribution. Fig.~\ref{fig:phasespace} shows the three phase distributions. Compared to the input probability density, dephasing broadens the distribution, while squeezing introduces two secondary peaks at $\pm \pi$: a border effect caused by the $\pi$-periodicity of the squeezing phase. Because of this latter effect the proper figure of merit for the phase uncertainty shall be the half width at half maximum (HWHM) of the central peak. 

Numerical evaluation of $p(\phi)$ in different regimes allows us to understand the role of the different parameters. We fix the OPO parameters $d$, $\eta_{\rm{in}}$, $\eta_{\rm{esc}}$ considering realistic values \cite{Cialdi} and consider several values for $\alpha$ and $\sigma$. Then, we compute and compare the HWHMs of the dephased distribution $\Gamma_D(\alpha, \sigma)$ and of the squeezed one $\Gamma_{\rm{out}}(\alpha, \sigma, d,\eta_{\rm{in}}, \eta_{\rm{esc}})$. 
As shown in Fig.~\ref{fig:Phasediag}, if the coherent signal $\alpha$ is small enough, $\Gamma_{\rm{out}}$ is constantly inferior than $\Gamma_D$, therefore the OPO proves to be always helpful regardless the values of $\sigma$. 
The reason is that even with zero noise, $\sigma=0$, there exists a regime where the squeezed HWHM $\Gamma_S(\alpha, d,\eta_{\rm{in}}, \eta_{\rm{esc}}) = \Gamma_{\rm{out}} (\alpha, \sigma=0, d,\eta_{\rm{in}}, \eta_{\rm{esc}})$ is smaller than the HWHM of the seed probability distribution $\Gamma_0(\alpha)$, since for small $\alpha$ the border peaks of the squeezed distribution centred in $\pm \pi$ induce a non-negligible reduction of the width of the central peak centred in $0$.
As a consequence, there exists a threshold coherent amplitude $\alpha_{\rm{th}}(d,\eta_{\rm{in}}, \eta_{\rm{esc}})$ such that if $\alpha < \alpha_{\rm{th}}(d,\eta_{\rm{in}}, \eta_{\rm{esc}})$ the OPO is always useful. The threshold is obtained by imposing the equality $\Gamma_0(\alpha_{\rm{th}})= \Gamma_S(\alpha_{\rm{th}}, d,\eta_{\rm{in}}, \eta_{\rm{esc}})$.
On the contrary, if $\alpha > \alpha_{\rm{th}}(d,\eta_{\rm{in}}, \eta_{\rm{esc}})$ we observe a $\sigma$-dependency. For small noise $\sigma$, we have $\Gamma_{\rm{out}} > \Gamma_D$ and so the OPO reveals useless, but on the contrary for large noise $\sigma$ the situation is reversed and $\Gamma_{\rm{out}} < \Gamma_D$. There exists a threshold noise $\sigma_{\rm{th}}(\alpha, d,\eta_{\rm{in}}, \eta_{\rm{esc}})$ such that the OPO keeps useful only for $\sigma > \sigma_{\rm{th}}(\alpha, d,\eta_{\rm{in}}, \eta_{\rm{esc}})$. Formally, such threshold noise is obtained by imposing that $\Gamma_D(\alpha, \sigma_{\rm{th}})= \Gamma_{\rm{out}} (\alpha, \sigma_{\rm{th}}, d,\eta_{\rm{in}}, \eta_{\rm{esc}})$.

\subsection{Indirect measurement of the phase}\label{sec:Indirect}
The second possible method involves two different and independent homodyne measurements of $q$ and $p$ to avoid the unavoidable excess noise induced by joint measurements. 
The value of the phase is then obtained as \cite{Cialdi}
\begin{equation}\label{eq:estimator}
\hat{\varphi}= \arctan\frac{\langle p \rangle}{\langle q \rangle}.
\end{equation} 
The exploitation of average values guarantees that squeezing border effects are absent and we choose the variance of $\hat{\varphi}$ as a good figure of merit. By exploiting the variance propagation law, we have
\begin{equation}
\Delta^2 \varphi = \frac{\langle q \rangle^2 \Delta^2 p + \langle p \rangle^2 \Delta^2 q}{\left(\langle q \rangle^2+ \langle p \rangle^2\right)^2}\,,
\end{equation}

The input state $\rho_0$ is Gaussian with
\begin{equation}
\langle q \rangle_0= \sqrt{2} \alpha,\quad \langle p \rangle_0=0, \quad \Delta^2q_0= \Delta^2p_0 = 1/2,
\end{equation}
therefore, $\Delta^2 \varphi_0 = 1/4 \alpha^2$. For the dephased state $\rho_D$ we have
\begin{equation}
\langle q \rangle_D= e^{-\sigma^2/2} \sqrt{2} \alpha, \quad \langle p \rangle_D=0, \quad
\end{equation}
and
\begin{equation}
\Delta^2q_D= \frac12 + \alpha^2 \left(1-e^{-\sigma^2}\right)^2, \quad
\Delta^2p_D = \frac12 + \alpha^2 \left(1-e^{-2\sigma^2}\right),
\end{equation}
leading to 
\begin{equation}
\Delta^2 \varphi_D= \frac{e^{\sigma^2}}{4\alpha^2} + \sinh \sigma^2 > \Delta^2 \varphi_0.
\end{equation}
Finally, the squeezed dephased state $\rho_{\rm{out}}$ has
\begin{equation}
\langle q \rangle_{\rm{out}}= e^{-\sigma^2/2} \sqrt{2} \tilde{\alpha}_q,\quad
\langle p \rangle_{\rm{out}}=0,\quad
\end{equation}
 and
 \begin{equation}
\Delta^2q_{\rm{out}}= \Sigma^2_q + \tilde{\alpha}^2_q \left( 1-e^{-\sigma^2}\right)^2,\quad
\Delta^2p_{\rm{out}}= \Sigma^2_p + \tilde{\alpha}^2_p \left( 1-e^{-2 \sigma^2}\right),
\end{equation}
thereafter, we obtain
\begin{equation}
\Delta^2 \varphi_{\rm{out}}= \frac{e^{\sigma^2}}{2\tilde{\alpha}^2_q} \Sigma^2_p + \frac{\tilde{\alpha}^2_p}{\tilde{\alpha}^2_q} \sinh \sigma^2.
\end{equation}
The variance of $\hat{\varphi}$ is given by two contributions, the former is correlated to the signal-to-noise ratio $\Delta^2 p/\langle q \rangle^2$ of the noiseless state $\rho_0$, the latter is a pure excess noise term. The excess noise term is always reduced after the OPO since $ \tilde{\alpha}_p / \tilde{\alpha}_q <1$. On the contrary, the signal-to-noise term after the OPO turns out to be lower than before only if $\Sigma^2_p/\tilde{\alpha}^2_q < 1/2\alpha^2$. This latter condition defines a threshold squeezing $d_{\rm{th}}$ at fixed values of $\eta_{\rm{in}}$ and $\eta_{\rm{esc}}$ via the equation
\begin{equation}
f(d_{\rm{th}}) \equiv \frac{ 4\eta_{\rm{in}}\eta_{\rm{esc}} }{ (1-d_{\rm{th}})^2 } \biggl[ 1 + \frac{d_{\rm{th}} }{\eta_{\rm{in}} } e^{-2r(d_{\rm{th}})}\biggr] -1 =0,
\end{equation}
and where $r(d)$ is defined in Eq.~(\ref{eq: squeezing_r}). If $d > d_{\rm{th}}(\eta_{\rm{in}}, \eta_{\rm{esc}})$ the OPO proves always useful, otherwise if $d < d_{\rm{th}}(\eta_{\rm{in}}, \eta_{\rm{esc}})$, the OPO amplifies the signal-to-noise ratio on quadratures but reduces the excess noise. For small $\sigma$ it is the signal-to-noise term to dominate, therefore the OPO is useless. When $\sigma$ is large the situation is reversed. There is a trade-off between the two contributions leading to the existence of a threshold noise $\sigma_{\rm{th}}$ above which squeezing shows a benefit. The threshold condition leads to
\begin{equation}\label{eq:Sigma_th}
\sigma^2_{\rm{th}}= \frac{1}{2} \ln \biggl[ \frac{2 \alpha^2 (\tilde{\alpha}^2_q-\tilde{\alpha}^2_p)}{\tilde{\alpha}^2_q+ 2 \alpha^2 (\tilde{\alpha}^2_q-\tilde{\alpha}^2_p - \Sigma^2_p)}\biggr],
\end{equation} 
and it agrees, at least qualitatively, to that obtained with a genuine phase measurement in the regime $\alpha \gg 1$ (see Fig.~\ref{fig:Phasediag}). We notice that this agreement is only \textit{qualitative}, since the two strategies presented here cannot be directly compared because they require a different number of measurements. In order to make a direct comparison between them one has to decrease $\alpha$ by a factor $1/\sqrt{2}$ in the indirect case to compensate the fact that two distinct measurements are performed on two copies of the state. Beside all, the agreement between the two strategies also provides a validation for the post-processing method of Sec. \ref{sec:Indirect}, which itself represents a convenient practical choice for experiments \cite{Cialdi}.

\section{Case II -- Estimation of a phase shift using a noisy probe}\label{sec: PhaseShift}
The second scenario in which we discuss the exploitation of an OPO is related to the estimation of a phase shift applied to a noisy probe. In particular, we analyze the protocol depicted in Fig.~\ref{fig: Cases}, case II, within the framework of quantum estimation theory (see Appendix~\ref{sec:QET}). We consider as probe the state
\begin{equation}\label{eq:probe}
\rho_{\rm{n}} = \mathcal{E}_{\rm{OPO}} \bigl(\mathcal{E}_\sigma (|\alpha\rangle \langle \alpha|) \bigr)
\end{equation}
on which we encode a phase shift $\theta$ through the unitary $U_\theta = e^{-i \theta a^\dagger a}$. After the encoding stage, the quantum state of radiation is sent into a channel until to reach a receiver, who performs measurements. In this case the task is to decide the optimal POVM $\{\Pi_x\}$ to infer the value of $\theta$. 
Here, we decide to restrict to a subclass of feasible measurements, i.e. \textit{homodyne measurements}. We investigate how the exploitation of the OPO modifies the quantum Fisher information (QFI) $H$ and the Fisher information (FI) $F$ of a homodyne detection of $x_\phi= \cos \phi \ q + \sin\phi \ p$. By comparing the two of them, we decide whether homodyne measurements can be optimal and, in particular, which is the best quadrature $x_{\phi_{\rm{max}}}$ that maximizes the FI. In the following, first of all we analyze the case in absence of phase diffusion as a benchmark and, secondly, we discuss the noisy case.

\subsection{Noiseless estimation scheme}\label{sec:noiselessproto}
In the absence of phase noise (but with the OPO still present), the probe state of the protocol in Fig.~\ref{fig: Cases}, case II, is the state $\rho_{\rm{OPO}}$ derived in Sec. \ref{sec:OPO}. Then, the encoded state reads
\begin{equation}
\rho_{\mathrm{nl},\theta} = U_\theta  \ \rho_{\rm{OPO}} \ U^\dagger_\theta.
\end{equation}
$\rho_{\rm{OPO}}$ being a Gaussian state with prime moments $ \textbf{R}_{\rm{OPO}} \equiv \langle \hat{\textbf{r}} \rangle= (\sqrt{2} \tilde{\alpha}_q, 0)$ and covariance $\boldsymbol \sigma_{\rm{OPO}} \equiv \langle \{(\hat{\textbf{r}}-\textbf{R}_{\rm{OPO}}),(\hat{\textbf{r}}-\textbf{R}_{\rm{OPO}})^T\}\rangle/2= \mathrm{Diag}[\Sigma^2_q, \Sigma^2_p]$, $\hat{\textbf{r}}= (q,p)$. Therefore, the statistical model $\rho_{\mathrm{nl},\theta}$ is still Gaussian with prime moments $\textbf{R}_\theta = \mathcal{R}_\theta \textbf{R}_{\rm{OPO}}$ and covariance $\boldsymbol\sigma_\theta = \mathcal{R}_\theta \boldsymbol \sigma_{\rm{OPO}}\mathcal{R}_\theta^T$, where $\mathcal{R}_\theta$ is the rotation matrix
\begin{equation}
\mathcal{R}_\theta = \begin{pmatrix} \cos\theta & \sin\theta \\ - \sin\theta & \cos\theta  \end{pmatrix}.
\end{equation}

In general, for a generic Gaussian state $\rho_\lambda$ with prime moments $\textbf{R}_\lambda$ and covariance $\boldsymbol \sigma_\lambda$, the QFI has the following analytical expression, 
\begin{equation}\label{eq:QFI-Gauss}
H(\lambda)= \frac{1}{2} \frac{\mathrm{Tr} \bigl[ (\boldsymbol\sigma_\lambda^{-1} \boldsymbol\sigma_\lambda')^2\bigr]}{1+\mu_\lambda^2}+2 \frac{\mu_\lambda'^2}{1-\mu_\lambda^4} + \textbf{R}_\lambda'^T \boldsymbol\sigma_\lambda^{-1} \textbf{R}_\lambda',
\end{equation}
where the $A' = \partial_\lambda A$ and $\mu_\lambda=(2\sqrt{\det\boldsymbol\sigma_\lambda})^{-1}$ is the purity of the Gaussian state $\rho_\lambda$ \cite{Serafini, Pinel, monras2013phase, PhysRevA.89.032128}.

\begin{figure}
\centering
\includegraphics[width=.6\textwidth]{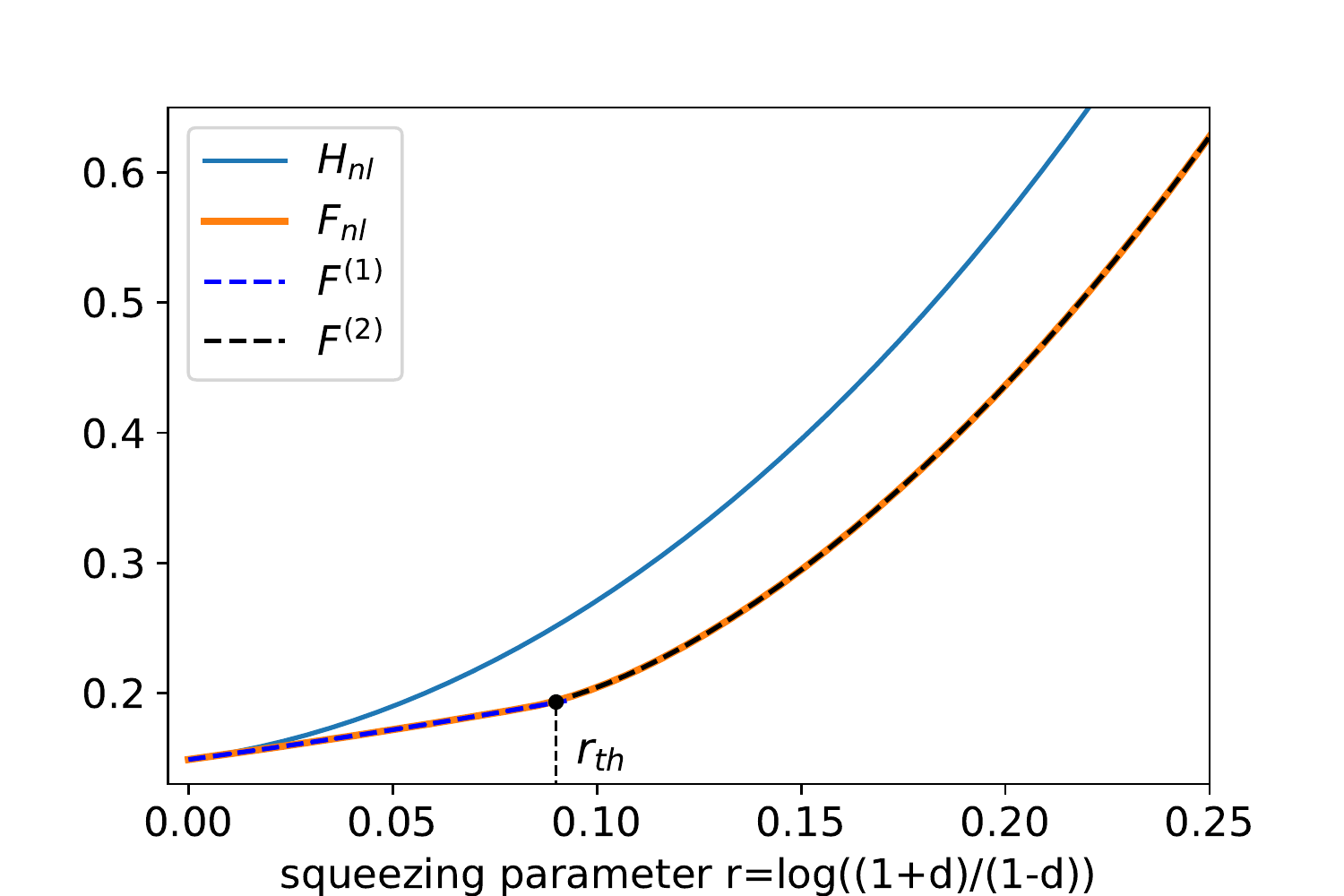}
\includegraphics[width=.6\textwidth]{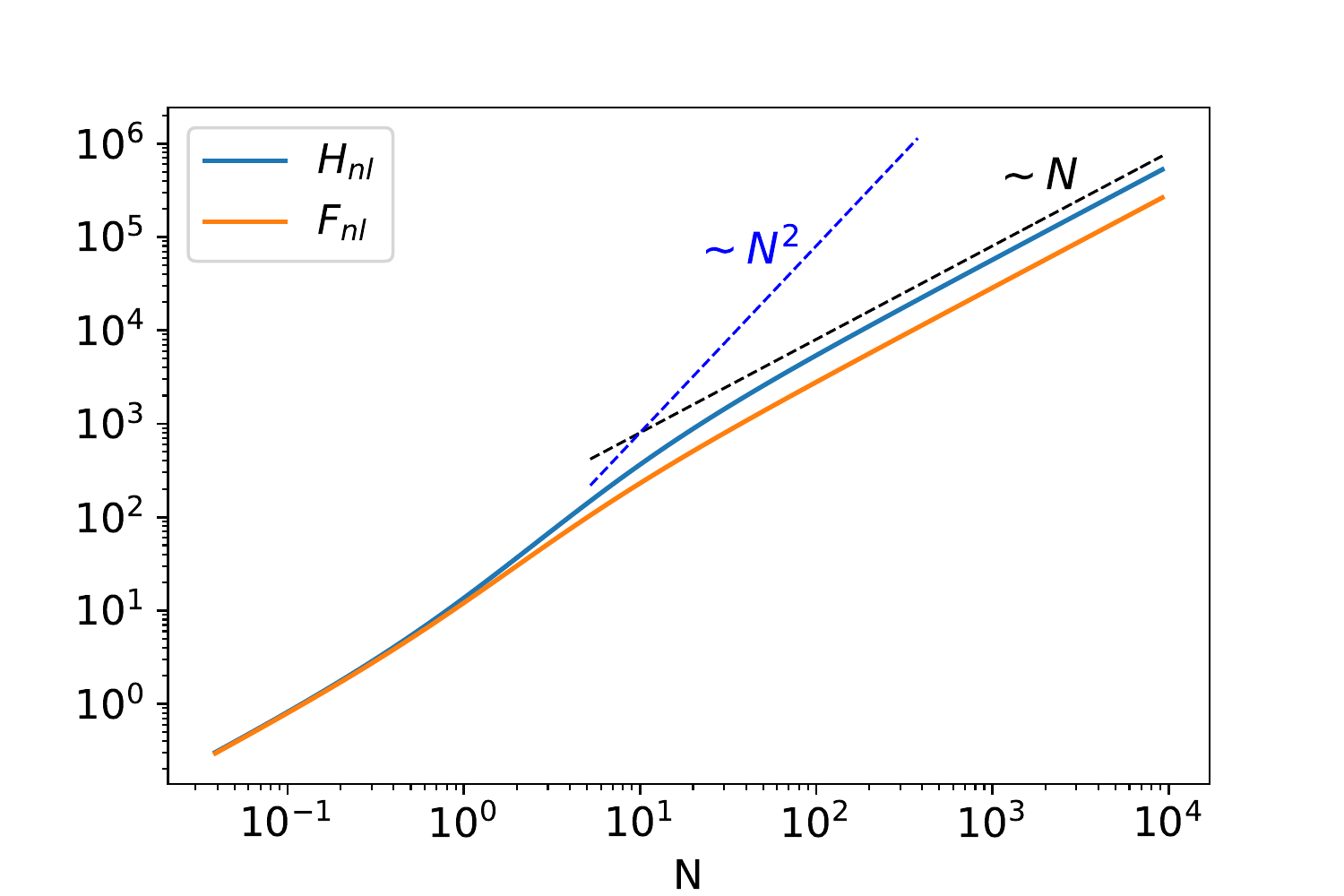}
\caption{Top: QFI and optimized FI as a function of $r$ for $\alpha=1$. For $r \geq r_{\rm{th}}(\alpha, \eta_{\rm{in}}, \eta_{\rm{esc}})$ the optimized quadrature is $\phi^{(2)}$ and $F_{\rm nl}$ coincides with $F^{(2)}$, whereas if $r<r_{\rm{th}}(\alpha, \eta_{\rm{in}}, \eta_{\rm{esc}})$ $F_{\rm nl}$ is equal to $F^{(1)}$ (see the text for details). Bottom: QFI and optimized FI as a function of $N$ for $\alpha=0.2$ (plot in log scale). The dashed lines refer to the Heisenberg scaling $\sim N^2$ and shot-noise one $\sim N$.  We used the realistic parameters $\eta_{\rm{in}}=0.01$, $\eta_{\rm{esc}}=0.93$.}\label{fig:QFI}
\end{figure}

In the very case in exam, Eq.~(\ref{eq:QFI-Gauss}) leads to 
\begin{equation}\label{eq:H}
H_{\rm{nl}} = 4\frac{(\Sigma^2_q-\Sigma^2_p)^2}{1+4\Sigma^2_p\Sigma^2_q} +2\frac{\tilde{\alpha}_q^2}{\Sigma^2_p},
\end{equation}
which is independent of $\theta$. The QFI depends on both $\alpha$ and $d$. However, the $\alpha$-dependence is only polynomial and so less relevant than the squeezing one. Therefore, in the following we keep $\alpha$ fixed and study the QFI dependence on the squeezing factor $r= \ln [(1+d)/(1-d)]$.
As depicted in Fig.~\ref{fig:QFI}, the QFI is a growing function on $r$. Thus, the most sensitive probe is obtained for $r\gg 1$, i.e. $d\approx 1$.
It is also worth to find out the asymptotic scaling of the QFI, by evaluating its dependence on the energy 
\begin{equation}\label{eq:N}
N= \mathrm{Tr}[\rho_{\mathrm{nl},\theta} \ a^\dagger a]= \frac{\Sigma^2_q+ \Sigma^2_p + \tilde{\alpha}^2_q -1}{2},
\end{equation} 
by keeping $\alpha$ fixed and varying only $d$.
Eventually, the QFI $H_{\rm{nl}}$ shows shot noise scaling:
\begin{equation}
H_{\rm{nl}} \approx \frac{4}{1-\eta_{\mathrm{esc}}} \frac{1+4\eta_{\rm{in}}\alpha^2}{1+2\eta_{\rm{in}}\alpha^2} \ N,
\end{equation}
whose origin can be addressed to the non-negligible losses characterizing the OPO dynamics.

As regards the analysis of the FI, it is helpful to write the probe state $\rho_{\rm{OPO}}$ in the form of a displayed squeezed thermal state \cite{OlivaresGauss, Ferraro2005}
\begin{equation}\label{eq:STS}
\rho_{\rm{OPO}} = D(\beta) \ S(\xi) \ \nu^{\rm th}(\bar{n}) \ S(\xi)^\dagger \ D(\beta)^\dagger,
\end{equation} 
where $D(\beta) = \exp(\beta a^\dagger - \beta^* a)$ is the displacement operator, $S(\xi) = \exp[\frac12\xi (a^{\dagger 2}-a^2)]$ is the squeezing operator and $\nu^{\rm{th}}(\bar{n})= \bar{n}^{a^\dagger a}/(\bar{n}+1)^{a^\dagger a+1}$ is a thermal state with mean number of photons $\bar{n}$. For the state $\rho_{\rm{OPO}}$ the values of the parameters are $ \beta= \tilde{\alpha}_q$ , $ \exp(2\xi)= \sqrt{\Sigma^2_q/\Sigma^2_p}$, $(1+2 \bar{n})^2 = 4 \Sigma^2_q \Sigma^2_p$.

Actually, it is proved that for a displayed squeezed thermal state the optimal measurement is not Gaussian \cite{Oh} and no homodyne can exactly reach the QFI. Nevertheless, it is still worth to construct an optimized homodyne measurement since it becomes nearly optimal (i.e. FI $\approx$ QFI) in the best working regime $r\gg 1$.

\begin{figure}
\centering
\includegraphics[width=.6\textwidth]{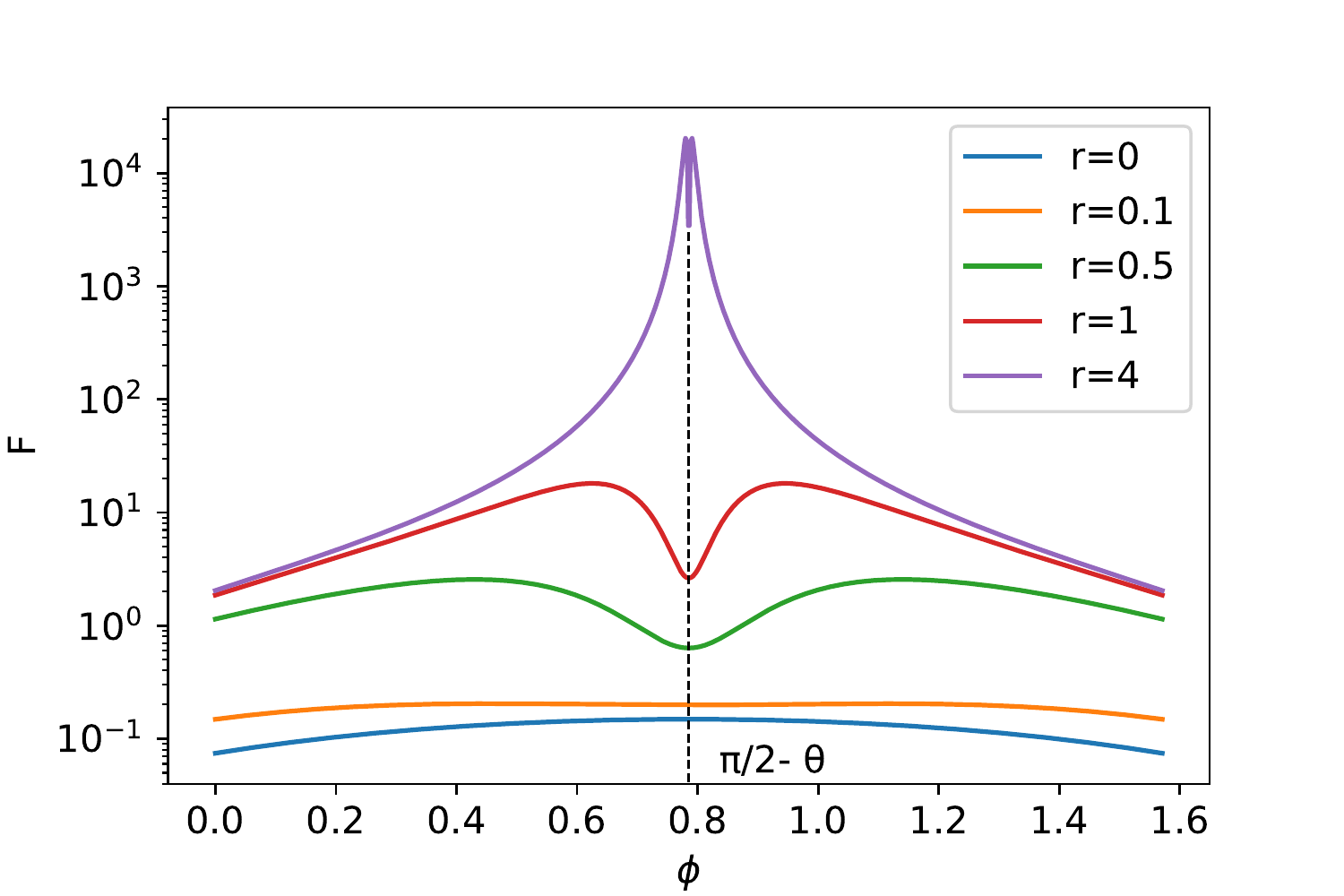}\\
\includegraphics[width=.6\textwidth]{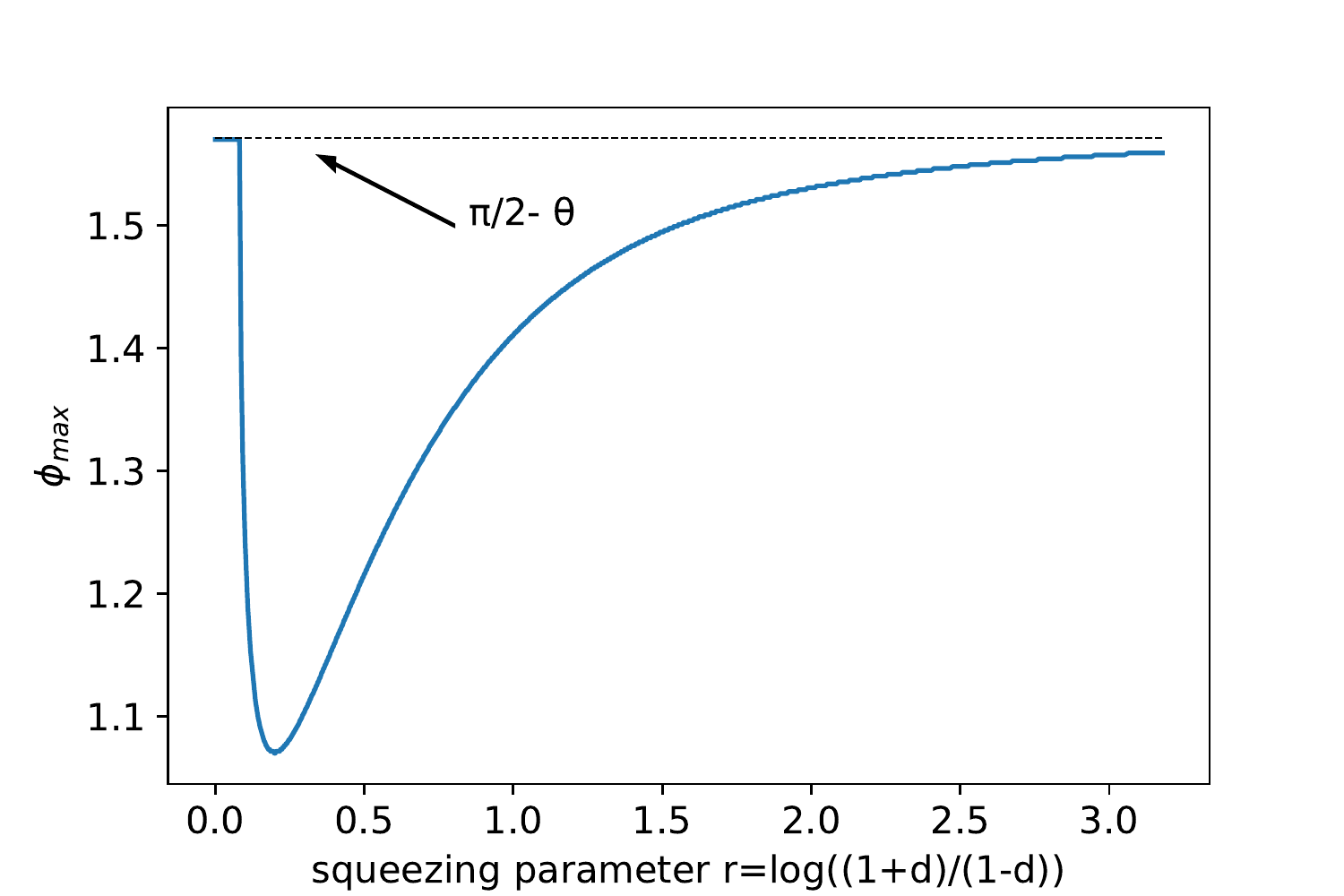}
\caption{Top: FI of the measurement of $x_\phi$ as a function of $\phi$ for different values of the squeezing parameter $r$.  Bottom: optimized quadrature $\phi_{\rm{max}}$ as a function of $r$. We set $\alpha=1$ and we used the realistic parameters $\eta_{\rm{in}}=0.01$, $\eta_{\rm{esc}}=0.93$.}\label{fig:OptimalPhi}
\end{figure}

To construct the optimized homodyne, we keep again $\alpha$ fixed and let only the squeezing parameter $r$ vary. For every $r$, we compute the FI associated with $x_\phi= \cos \phi \ q + \sin\phi \ p$ as a function of $\phi$ and find the value $\phi_{\rm{max}}$ maximizing it, as depicted in Fig.~\ref{fig:OptimalPhi}. In the end, the optimized FI $F_{\rm{nl}}$, displayed in Fig. \ref{fig:QFI}, turns out to be a piecewise-defined function, since there exist two quadratures candidate to the choice of $\phi_{\rm{max}}$  \cite{Oh}:
\begin{enumerate}
\item $\phi^{(1)}=\pi/2-\theta$ whose corresponding FI is
\begin{equation}\label{eq: F_opt 1}
F^{(1)}= \frac{2 \tilde{\alpha}_q^2}{\Sigma^2_p},
\end{equation}
\item $\phi^{(2)}=\pi/2-\theta-\chi/2$, where $\chi$ satisfies
\begin{equation}
\cos\chi = \frac{(\Sigma^2_q-\Sigma^2_p)^3 +\Sigma^2_q \tilde{\alpha}_q^2 (\Sigma^2_q+\Sigma^2_p)}{(\Sigma^2_q+\Sigma^2_p)(\Sigma^2_q-\Sigma^2_p)^2 +\Sigma^2_q \tilde{\alpha}_q^2 (\Sigma^2_q-\Sigma^2_p)}
 \quad (d \neq 0),
\end{equation}
for which the corresponding value of the FI reads 
\begin{equation}\label{eq: F_opt 2}
F^{(2)}= \frac{\bigl[(\Sigma^2_q-\Sigma^2_p)^2 + \Sigma^2_q \tilde{\alpha}_q^2\bigr]^2}{2 \Sigma^2_q \Sigma^2_p (\Sigma^2_q-\Sigma^2_p)^2}.
\end{equation}
\end{enumerate}
The second measurement is well defined only if $|\cos \chi|\leq 1$, namely
\begin{equation}\label{eq:Cond}
\tilde{\alpha}_q \leq (\Sigma^2_q-\Sigma^2_p)/\sqrt{\Sigma^2_q},
\end{equation} 
providing a threshold squeezing $r_{\rm{th}}(\alpha, \eta_{\rm{in}}, \eta_{\rm{esc}})$ such that for smaller $r$ the function $F^{(2)}$ has no physical meaning and the optimized quadrature is $\phi^{(1)}$, while for larger $r$ the homodyne measurement of quadrature $\phi^{(2)}$ is well defined and the optimized quadrature becomes $\phi^{(2)}$. The physical explanation of such behaviour becomes clear by observing the $\phi$ dependence of the FI displayed in Fig. \ref{fig:OptimalPhi}. For small $r$, the function has a single maximum at $\pi/2-\theta$, but for larger $r$ the maximum splits into two symmetric peakes and $\pi/2-\theta$ turns into a local minimum. Then, by increasing $r$ further the position of the peakes asymptotically converges to $\pi/2-\theta$.
As regards the energy scaling of $F_{\rm{nl}}$, by rearranging Eq.~(\ref{eq: F_opt 2}) we get again shot noise scaling
\begin{equation}
F_{\rm{nl}} \approx 2 \frac{1+2\eta_{in}\alpha^2}{1- \eta_{esc}} \ N.
\end{equation}

\subsection{Noisy estimation scheme}
\begin{figure}
\centering
\includegraphics[width=.6\textwidth]{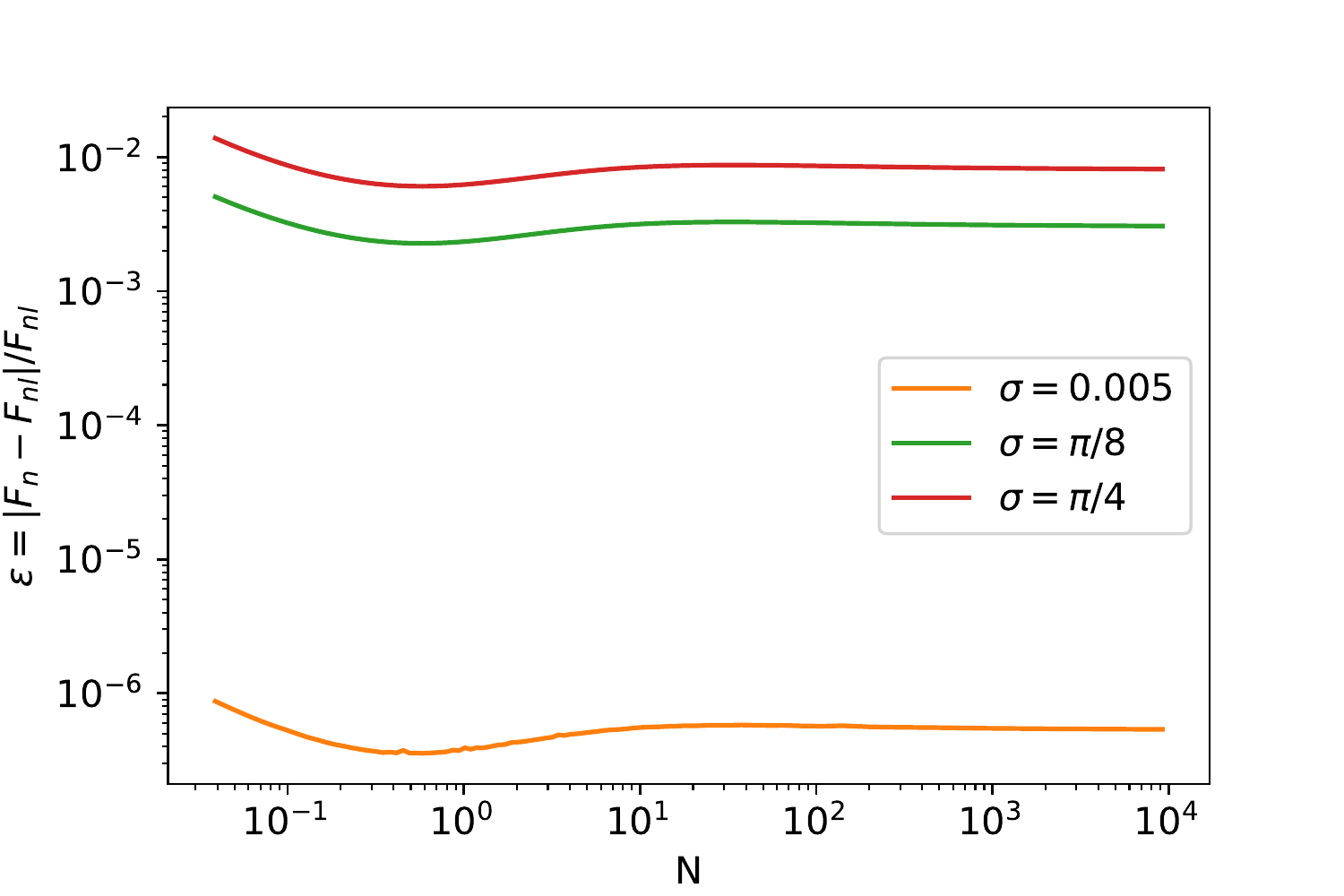} \\
\includegraphics[width=.6\textwidth]{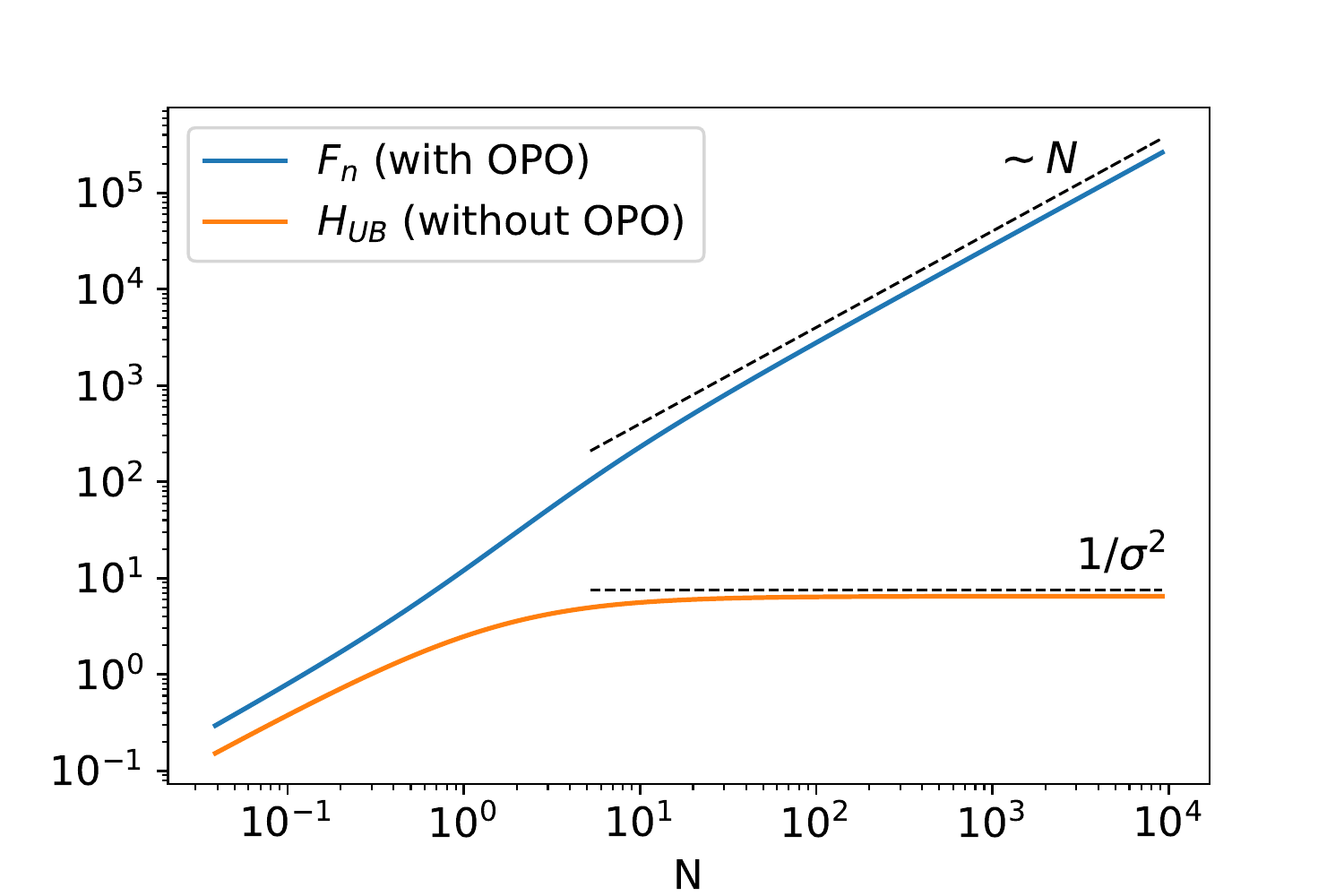}
\caption{Top: Relative fluctuations $\epsilon$ given in Eq.~(\ref{rel:fluct}) as a function of $N$ for different values of the noise parameter $\sigma$. Bottom: Plot of the $F_{\rm{n}}$ and $H_{\rm{UB}}$ as functions of $N$ for $\sigma=\pi/8$. We set $\theta=0$, $\alpha=0.2$ and we used the realistic parameters $\eta_{\rm{in}}=0.01$, $\eta_{\rm{esc}}=0.93$.}\label{fig:Fn}
\end{figure}

If phase noise is present, the statistical model reads
\begin{equation}
\rho_{\mathrm{n},\theta} = U_\theta  \ \rho_{\rm{n}} \ U^\dagger_\theta,
\end{equation}
where $\rho_{\rm{n}}$ is still given in Eq.~(\ref{eq:probe}).
The presence of the this kind of noise prevents to obtain analytical solutions. Therefore, we keep the results of Sec. \ref{sec:noiselessproto} as a benchmark and consider how the noise affects a specific case. We perform a homodyne of $x_{\phi_{\rm{max}}}$, where $\phi_{\rm{max}}$ is the optimized quadrature of Fig.~\ref{fig:OptimalPhi}. For several values of $\sigma$, we analyze the dependence of the FI 
\begin{equation}
F_{\rm{n}} = \int dx \frac{\bigl[\partial_\theta p(x|\theta)\bigr]^2}{p(x|\theta)}, \qquad p(x|\theta) = \mathrm{Tr}[\rho_{\mathrm{n},\theta} \ x_{\phi_{\rm{max}}}],
\end{equation}
on the energy $N$ of Eq.~(\ref{eq:N}). Moreover, to compare the noiseless and noisy cases we introduce the relative fluctuations parameter
\begin{equation}\label{rel:fluct}
\epsilon = \frac{|F_{\rm{n}}-F_{\rm{nl}}|}{F_{\rm{nl}}}.
\end{equation}
The numerical results of $\epsilon$, depicted in Fig.~\ref{fig:Fn}, show that the exploitation of the OPO is able to compensate almost completely the detriments of phase noise. 
In particular, using an OPO is crucial to maintain the shot noise regime for all values of $\sigma$.
Otherwise, if we consider the protocol of Fig.~\ref{fig: Cases}, case II, without the OPO, there exists an upper bound for the QFI \cite{Escher}, namely
\begin{equation}
H_{\rm{without \ OPO}} \leq H_{\rm{UB}} = \frac{4 N}{1+4N \sigma^2},
\end{equation}
saturating for large $N$ to the value $1/\sigma^2$.

\section{Conclusions}\label{s:concl}
In summary, we have studied several scenarios where phase noise prevents the use of optical phase as a degree of freedom for quantum information tasks and discussed whether an OPO may be employed to mitigate, or even compensate, the effects of noise.
Such a method could be interesting with the intent of reducing the noise of the laser in a regime of high frequency.

At first, we have developed a block-diagram model to describe an OPO in the form of a subsequent application of Gaussian operations: beam splitters, phase-sensitive amplification, phase-sensitive phase shift and squeezing. Such description in the Schr\"odinger picture allows to give an explicit expression for the output state of radiation. Indeed, given a initial coherent state the output state is a Gaussian state with well-defined prime moments and covariance.

With the new description of the OPO we have addressed the first case under investigation: the measurement of the phase of a quantum state of radiation. We have introduced two possible approaches. A first standard approach involves the introduction of a phase POVM (implemented through heterodyne detection). It leads to the conclusion that the OPO reduces phase noise, i.e. the width of the probability distribution associated with the POVM, for small signal amplitude, or large signal amplitude and large dephasing.
The second approach consists in a post-processing method based on the outcomes of two distinct homodyne detections and brings the same phenomenology as before in the regime of large coherent amplitudes.
In both cases, according to the value of parameters $\alpha, \sigma, d$, there exists a regime where the procedure gives a phase outcome with a larger uncertainty than the one of the initial coherent state and another regime in which the uncertainty is smaller. This leads to the conclusion that in such regime the OPO is able to \textit{fully}, or at least \textit{partially}, compensate the noise.

The second scenario discussed has been an estimation scheme based on the encoding of a phase shift $\theta$ on the probe state. We have considered a dephased coherent state passed through an OPO as probe state and searched for the optimal POVM to detect $\theta$ within the subclass of homodyne measurements. We have studied in detail the noiseless protocol, obtaining the optimized quadrature $\phi_{\rm{max}}$ that allows to infer $\theta$ with negligible uncertainty. Passing to the noisy case, such situation is still maintained. The noise affects weakly the FI and shot noise scaling, i. e. the proper scaling of the input coherent state, is conserved. Therefore, in such case the noise mitigation by the OPO is \textit{complete} as regards the scaling with the probe average photon number.

Our results confirm that it is possible to develop suitable OPO-based strategies to compensate phase noise with current technology and, thus, pave the way for the full exploitation of optical phase in quantum technologies.

\section*{Acknowledgements}
This work has been supported by MAECI, Project No.~PGR06314 ``ENYGMA'' and by University of Milan, Project No.~\mbox{RV-PSR-SOE-2020-SOLIV} ``\mbox{S-O}~PhoQuLis''

\appendix

\section{Elements of quantum estimation theory}\label{sec:QET}
The estimation of a parameter is a frequent task in quantum mechanics, since several physical quantities cannot be directly measured. Here we present the basic features of the theory behind it \cite{Paris-2009, Helstrom}. We consider a family of quantum states labelled by a parameter $\lambda$, $\{\rho_\lambda\}_\lambda$, usually called \textit{statistical model}. Usually, we perform a generalized measurement described by a positive operator-valued measure (POVM) $\{\Pi_x\}$, obtaining a statistical sample of $M$ outcomes $\textbf{x}=\{x_1, ..., x_M\}$. This sample is processed by means of a map $\hat{\lambda}(\textbf{x})$, called an \textit{estimator}, to infer the value of the parameter $\lambda$. The task is to find the optimal POVM that allows to estimate the value of $\lambda$ with the lowest possible uncertainty, i.e., the maximum precision.
The conditional probability of the outcome $x$ given $\lambda$ is 
\begin{equation}
p(x|\lambda) = \mathrm{Tr} [\rho_\lambda \Pi_x].
\end{equation}
If the estimator is unbiased, there exists a lower bound to its variance, depending on the \textit{Fisher information} (FI) of the distribution $p(x|\lambda)$ 
\begin{equation}
F(\lambda) = \int dx \frac{\bigl[\partial_\lambda p(x|\lambda)\bigr]^2}{p(x|\lambda)}.
\end{equation}
The bound is the so called \textit{Cramér-Rao bound} and reads
\begin{equation}\label{eq: Cramerrao}
\text{Var} [\hat{\lambda}] \geq \frac{1}{M F(\lambda)},
\end{equation}
where we introduced the variance $\text{Var} [\hat{\lambda}]  = \text{E}[\hat{\lambda}^2]-\text{E}[\hat{\lambda}]^2$ with
\begin{equation}
\text{E} [\hat{\lambda}^k] = \int dx\, p(x|\lambda)\, \hat{\lambda}(x)^k,\quad k \in {\mathbb N}.
\end{equation}
However, a more strict bound, independent of the particular measurement performed, may be obtained \cite{Malley, Braunstein1994, Braunstein1996, Brody1998}. We define the Symmetric Logarithmic Derivative (SLD) $L_\lambda$ by the Ljapunov equation $2 \partial_\lambda \rho_\lambda = L_\lambda \rho_\lambda+ \rho_\lambda L_\lambda$ and the \textit{Quantum Fisher Information} (QFI) as \cite{Helstrom}
\begin{equation}
H(\lambda) = \mathrm{Tr} [\rho_\lambda L^2_\lambda].
\end{equation}
The QFI leads to the \textit{Quantum Cramér-Rao bound}
\begin{equation}\label{eq: QCramerrao}
\text{Var} [\hat{\lambda}] \geq \frac{1}{M H(\lambda)}.
\end{equation}
\\
\indent In assessing a quantum estimation scheme, both the QFI and the FI are important tools. The QFI identifies the ultimate limits on precision allowed by quantum mechanics, independent of the measurement, while the FI fixes the minimum possible uncertainty given a particular measurement strategy, namely, a POVM. In this present work we will address a subclass of possible measurements, that is homodyne measurements, and determine their performance by comparing the QFI and FI.


\begin{thebibliography}{}
\newcommand{\enquote}[1]{``#1''}

\end{thebibliography}


\begin{thebibliography}{10}
\newcommand{\enquote}[1]{``#1''}

\bibitem{Caves1981}
C.~M. Caves, \enquote{Quantum-mechanical noise in an interferometer,}
  {\protect\JournalTitle{Physical Review D}} \textbf{23}, 1693--1708 (1981).

\bibitem{demkowicz2009}
R.~Demkowicz-Dobrza\'nski, U.~Dorner, B.~J. Smith, J.~S. Lundeen, W.~Wasilewski,
  K.~Banaszek, and I.~A. Walmsley, \enquote{Quantum phase estimation with lossy
  interferometers,} {\protect\JournalTitle{Physical Review A}} \textbf{80},
  013825 (2009).

\bibitem{sparaciari2015}
C.~Sparaciari, S.~Olivares, and M.~G.~A. Paris, \enquote{Bounds to precision
  for quantum interferometry with Gaussian states and operations,}
  {\protect\JournalTitle{J. Opt. Soc. Am. B}} \textbf{32}, 1354--1359 (2015).

\bibitem{Kazovsky2006}
L.~G. Kazovsky, G.~Kalogerakis, and W.~Shaw, \enquote{Homodyne
  phase-shift-keying systems: Past challenges and future opportunities,}
  {\protect\JournalTitle{Journal of Lightwave Technology}} \textbf{24},
  4876--4884 (2006).

\bibitem{Olivares2013}
S.~Olivares, S.~Cialdi, F.~Castelli, and M.~G.~A. Paris, \enquote{Homodyne
  detection as a near-optimum receiver for phase-shift-keyed binary
  communication in the presence of phase diffusion,}
  {\protect\JournalTitle{Physical Review A}} \textbf{87}, 050303 (2013).

\bibitem{mondin15}
M.~Mondin, F.~Daneshgaran, I.~Bari, M.~T. Delgado, S.~Olivares, and M.~G.~A.
  Paris, \enquote{Soft-metric-based channel decoding for photon counting
  receivers,} {\protect\JournalTitle{IEEE Journal of Selected Topics in Quantum
  Electronics}} \textbf{21}, 62--68 (2015).

\bibitem{Susskind}
L.~Susskind and J.~Glogower, \enquote{Quantum mechanical phase and time
  operator,} {\protect\JournalTitle{Physics Physique Fizika}} \textbf{1}, 49
  (1964).

\bibitem{Louisell}
W.~H. Louisell, \enquote{Amplitude and phase uncertainty relations,}
  {\protect\JournalTitle{Physics Letters}} \textbf{7}, 60--61 (1963).

\bibitem{Paris-1994}
G.~M. D’Ariano and M.~G.~A. Paris, \enquote{Lower bounds on phase sensitivity
  in ideal and feasible measurements,} {\protect\JournalTitle{Physical Review
  A}} \textbf{49}, 3022--3036 (1994).

\bibitem{Lalovic}
D.~I. Lalović, D.~M. Davidović, and A.~R. Tan\v{c}ić, \enquote{Quantum phase
  from the glauber model of linear phase amplifiers,}
  {\protect\JournalTitle{Physical Review Letters}} \textbf{81}, 1223--1226
  (1998).

\bibitem{Ezra2008}
E.~Ip, A.~P.~T. Lau, D.~J.~F. Barros, and J.~M. Kahn, \enquote{Coherent
  detection in optical fiber systems,} {\protect\JournalTitle{Opt. Express}}
  \textbf{16}, 753--791 (2008).

\bibitem{Qubit2}
D.~Brivio, S.~Cialdi, S.~Vezzoli, B.~T. Gebrehiwot, M.~G. Genoni, S.~Olivares,
  and M.~G.~A. Paris, \enquote{Experimental estimation of one-parameter qubit
  gates in the presence of phase diffusion,} {\protect\JournalTitle{Physical
  Review A}} \textbf{81}, 012305 (2010).

\bibitem{Qubit1}
B.~Teklu, M.~G. Genoni, S.~Olivares, and M.~G.~A. Paris, \enquote{Phase
  estimation in the presence of phase diffusion: the qubit case,}
  {\protect\JournalTitle{Physica Scripta}} \textbf{2010}, 014062 (2010).

\bibitem{Genoni}
M.~G. Genoni, S.~Olivares, and M.~G.~A. Paris, \enquote{Optical phase
  estimation in the presence of phase diffusion,}
  {\protect\JournalTitle{Physical Review Letters}} \textbf{106}, 153603 (2011).

\bibitem{Genoni2012}
M.~G. Genoni, S.~Olivares, D.~Brivio, S.~Cialdi, D.~Cipriani, A.~Santamato,
  S.~Vezzoli, and M.~G.~A. Paris, \enquote{Optical interferometry in the
  presence of large phase diffusion,} {\protect\JournalTitle{Phys. Rev. A}}
  \textbf{85}, 043817 (2012).

\bibitem{Trapani2015}
J.~Trapani, B.~Teklu, S.~Olivares, and M.~G.~A. Paris, \enquote{Quantum phase
  communication channels in the presence of static and dynamical phase
  diffusion,} {\protect\JournalTitle{Phys. Rev. A}} \textbf{92}, 012317 (2015).

\bibitem{Jarzyna2016}
M.~Jarzyna, V.~Lipi\'{n}ska, A.~Klimek, K.~Banaszek, and M.~G.~A. Paris,
  \enquote{Phase noise in collective binary phase shift keying with hadamard
  words,} {\protect\JournalTitle{Opt. Express}} \textbf{24}, 1693--1698 (2016).

\bibitem{Bina}
M.~Bina, A.~Allevi, M.~Bondani, and S.~Olivares, \enquote{Phase-reference
  monitoring in coherent-state discrimination assisted by a photon-number
  resolving detector,} {\protect\JournalTitle{Scientific Reports}} \textbf{6},
  1--9 (2016).

\bibitem{DiMario2019}
M.~T. DiMario, L.~Kunz, K.~Banaszek, and F.~E. Becerra, \enquote{Optimized
  communication strategies with binary coherent states over phase noise
  channels,} {\protect\JournalTitle{npj Quantum Information}} \textbf{5}, 65
  (2019).

\bibitem{Bose1}
I.~Tikhonenkov, M.~G. Moore, and A.~Vardi, \enquote{Optimal Gaussian squeezed
  states for atom interferometry in the presence of phase diffusion,}
  {\protect\JournalTitle{Physical Review A}} \textbf{82}, 043624 (2010).

\bibitem{Bose2}
Y.~C. Liu, G.~R. Jin, and L.~You, \enquote{Quantum-limited metrology in the
  presence of collisional dephasing,} {\protect\JournalTitle{Physical Review
  A}} \textbf{82}, 045601 (2010).

\bibitem{Josephson}
G.~Ferrini, D.~Spehner, A.~Minguzzi, and F.~W.~J. Hekking, \enquote{Noise in
  Bose Josephson junctions: Decoherence and phase relaxation,}
  {\protect\JournalTitle{Physical Review A}} \textbf{82}, 033621 (2010).

\bibitem{Cialdi}
S.~Cialdi, E.~Suerra, S.~Olivares, S.~Capra, and M.~G.~A. Paris,
  \enquote{Squeezing phase diffusion,} {\protect\JournalTitle{Physical Review
  Letters}} \textbf{124}, 163601 (2020).

\bibitem{Carrara}
G.~Carrara, M.~G. Genoni, S.~Cialdi, M.~G.~A. Paris, and S.~Olivares,
  \enquote{Squeezing as a resource to counteract phase diffusion in optical
  phase estimation,} {\protect\JournalTitle{Physical Review A}} \textbf{102},
  062610 (2020).

\bibitem{aguideto}
H.-A. Bachor and T.~C. Ralph, \emph{A Guide to Experiments in Quantum Optics},
  Physics textbook (Wiley-VCH, 2004).

\bibitem{OlivaresGauss}
S.~Olivares, \enquote{Quantum optics in the phase space,}
  {\protect\JournalTitle{The European Physical Journal Special Topics}}
  \textbf{203}, 3--24 (2012).

\bibitem{PhysRevA.54.4495}
G.~M. D'Ariano, M.~G.~A. Paris, and R.~Seno, \enquote{Feedback-assisted
  homodyne detection of phase shifts,} {\protect\JournalTitle{Phys. Rev. A}}
  \textbf{54}, 4495--4504 (1996).

\bibitem{PhysRevLett.67.1426}
J.~W. Noh, A.~Foug\`eres, and L.~Mandel, \enquote{Measurement of the quantum
  phase by photon counting,} {\protect\JournalTitle{Phys. Rev. Lett.}}
  \textbf{67}, 1426--1429 (1991).

\bibitem{Serafini}
A.~Serafini, \emph{Quantum Continuous Variables: A Primer of Theoretical
  Methods} (CRC Press, Taylor \& Francis Group, 2017).

\bibitem{Pinel}
O.~Pinel, P.~Jian, N.~Treps, C.~Fabre, and D.~Braun, \enquote{Quantum parameter
  estimation using general single-mode Gaussian states,}
  {\protect\JournalTitle{Physical Review A}} \textbf{88}, 040102 (2013).

\bibitem{monras2013phase}
A.~Monras, \enquote{Phase space formalism for quantum estimation of Gaussian
  states,}  eprint arXiv:1303.3682 (2013).

\bibitem{PhysRevA.89.032128}
Z.~Jiang, \enquote{Quantum fisher information for states in exponential form,}
  {\protect\JournalTitle{Phys. Rev. A}} \textbf{89}, 032128 (2014).

\bibitem{Ferraro2005}
A.~Ferraro, S.~Olivares, and M.~G.~A. Paris, \emph{Gaussian states in quantum
  information} (Bibliopolis Napoli, 2005).

\bibitem{Oh}
C.~Oh, C.~Lee, C.~Rockstuhl, H.~Jeong, J.~Kim, H.~Nha, and S.-Y. Lee,
  \enquote{Optimal Gaussian measurements for phase estimation in single-mode
  Gaussian metrology,} {\protect\JournalTitle{npj Quantum Information}}
 \textbf{5}, 10 (2019).

\bibitem{Escher}
B.~M. Escher, L.~Davidovich, N.~Zagury, and R.~L. de~Matos~Filho,
  \enquote{Quantum metrological limits via a variational approach,}
  {\protect\JournalTitle{Phys. Rev. Lett.}} \textbf{109}, 190404 (2012).
  
\bibitem{Paris-2009}
M.~G.~A. Paris, \enquote{Quantum estimation for quantum technology,}
  {\protect\JournalTitle{International Journal of Quantum Information}}
  \textbf{7}, 125--137 (2009).
  
\bibitem{Helstrom}
C.~W. Helstrom, \emph{Quantum Detection and Estimation Theory}, Mathematics in
  Science and Engineering 123 (Elsevier, Academic Press, 1976).

\bibitem{Malley}
J.~D. Malley and J.~Hornstein, \enquote{Quantum statistical inference,}
  {\protect\JournalTitle{Statistical Science}} pp. 433--457 (1993).

\bibitem{Braunstein1994}
S.~L. Braunstein and C.~M. Caves, \enquote{Statistical distance and the
  geometry of quantum states,} {\protect\JournalTitle{Physical Review Letters}}
  \textbf{72}, 3439 (1994).

\bibitem{Braunstein1996}
S.~L. Braunstein, C.~M. Caves, and G.~J. Milburn, \enquote{Generalized
  uncertainty relations: theory, examples, and lorentz invariance,}
  {\protect\JournalTitle{Annals of Physics}} \textbf{247}, 135--173 (1996).

\bibitem{Brody1998}
D.~C. Brody and L.~P. Hughston, \enquote{Statistical geometry in quantum
  mechanics,} {\protect\JournalTitle{Proceedings of the Royal Society of
  London. Series A: Mathematical, Physical and Engineering Sciences}}
  \textbf{454}, 2445--2475 (1998).

\end{thebibliography}

\end{document}